\documentclass[12pt]{article}
\usepackage{amsmath}
\usepackage{epsfig}
\title{Hadron Polarizabilities}
\author{Barry R. Holstein\\
Department of Physics-LGRT\\
University of Massachusetts\\
Amherst, MA  01003 USA\\
and\\
Stefan Scherer\\
PRISMA Center of Excellence\\
Institut f\"{u}r Kernphysik\\
Johannes Gutenberg-Universit\"{a}t Mainz\\
D-55099 Mainz, Germany
}
\date{December 31, 2013}
\begin{document}
\begin{titlepage}
\maketitle
\begin{abstract}
   Electromagnetic polarizabilities describe the response of a system to the application of
an external quasi-static electric or magnetic field.
In this article experimental and theoretical work addressing the
polarizabilities of the light hadrons is examined.
\end{abstract}
\end{titlepage}

\section{Introduction}

   The concept of polarizability is well known from classical electrodynamics, where the electric and magnetic
polarizabilities $\alpha_E,\beta_M$ are simply the coefficients of proportionality between applied quasi-static
electric/magnetizing fields and the resultant {\it induced} electric/magnetic dipole moments \cite{jdj}.
   As discussed below, this concept is also applicable to hadronic systems, where the applied fields are provided by
electromagnetic radiation and the measurements are performed via Compton scattering
\cite{Klein:1955zz,Friar:1988rr,Holstein:1990qy,Holstein:1992xr,Lvov:1993fp,Scherer:1999yw,
Drechsel:2002ar,Schumacher:2005an,Downie:2011mm}.
   In this regard, polarizabilities provide the leading-order correction to the well-known Thomson
scattering amplitude and a series of experiments have measured $\alpha_E,\beta_M$ for various hadronic systems
such as the proton \cite{Federspiel:1991yd,Zieger:1992jq,Mac,Bla,Olm}, the neutron
\cite{Schmiedmayer:1991zz,Rose:1990zz,Kossert:2002ws,Lundin:2002jy}, and even the charged pion
\cite{Antipov:1982kz,Antipov:1984ez,Aibergenov:1986gi,Babusci:1991sk,Ahrens:2004mg,Friedrich:2013iya}.
   Such quantities provide valuable information concerning hadronic structure and are fundamental
properties \cite{Holstein:1990qy,Holstein:1992xr} which are now included by the particle data group
in their listings \cite{pdg}.
   At the same time, they are also discussed in the context of precision determinations of
other physical observables such as the Lamb shift in muonic hydrogen \cite{Nevado:2007dd,Birse:2012eb}
and the hadronic light-by-light contribution to the muon anomalous magnetic moment \cite{Engel:2012xb}.
   At higher order there exist additional {\it spin-dependent}
and {\it spin-independent} polarizabilities which provide new structure probes and are just now beginning to be
measured experimentally \cite{Ragusa:1993rm,Ragusa:1994pm,Babusci:1998ww,Holstein:1999uu}.
   Information about the distribution of polarizability structure within the
proton has been provided by virtual Compton scattering measurements
\cite{Roche:2000ng,Laveissiere:2004nf,Bourgeois:2006js,Janssens:2008qe,Bourgeois:2011zz,Fonvieille:2012cd}.
   On the theoretical side there exists a great deal of work in the literature involving the analysis
of Compton scattering, both real and virtual, in terms of polarizabilities, using hadronic models as well
as with more rigorous methods such as the use of dispersion theory \cite{lvo,pas} and/or effective field
theory (chiral perturbation theory) \cite{Gasser:1983yg,Gasser:1987rb}.

   The purpose of this article is to provide an overview of the existing body of
theoretical and experimental probes of electromagnetic polarizabilities as well as to provide
a blueprint for possible future work.
   In Sec.~2, we introduce the basic concept of electric and magnetic (scalar, {\it i.e.}, spin-independent)
polarizabilities and examine their experimental determination as well as various theoretical
approaches to these quantities.
   In Sec.~3, we examine the corresponding spin-dependent (vector) polarizabilities which arise at one
higher order in the Compton amplitude, discussing the theoretical predictions for such quantities as
well as ongoing experimental programs which seek to measure them.
   In Sec.~4, we look at even higher-order polarizabilities and compare theoretical predictions
to results obtained from dispersive approaches to Compton scattering.
   In Sec.~5, we discuss how virtual Compton scattering has been used in order to determine
polarizability distributions within the proton.
  Finally, in a concluding Section 6, we present a brief summary of what has been learned as well
as a glimpse into possible future work in this area.

\section{Scalar Polarizabilities}
\subsection{Macroscopic Systems}
  In classical physics when elementary systems having structure are probed via
application of an external electric field, various responses are conceivable.
  In the simplest case, the applied field leads to a charge separation and
thereby to an induced electric dipole moment, with positive (negative)
charges moving in the direction (opposite to the direction) of the field
\cite{jdj}.
   In the case of orientation polarization, preexisting but initially randomly
oriented permanent dipole moments are lined up by the applied field.
   In anisotropic materials, even though the polarization still depends linearly
on the field, its direction is not necessarily parallel to the applied field but
is determined in terms of a polarizability tensor of second rank \cite{bd}.
   Finally, in ferroelectric materials the polarization is a nonlinear function
of the electric field showing, in addition, a hysteresis effect.

   In the first of the above cases, the electric polarizability $\alpha_E$ of a
system is simply the constant of proportionality between the applied static and
uniform field $\mathbf{E}$ and the induced electric dipole moment $\mathbf{p}$,
\begin{equation}
\mathbf{p}=4\pi\alpha_E\mathbf{E},
\end{equation}
corresponding to a potential energy
\begin{equation}
U_E=-\frac{1}{2}4\pi\alpha_E\mathbf{E}^2,
\quad\textnormal{\it i.e.},\quad
\mathbf{p}=-\frac{\partial U_E}{\partial \mathbf{E}}.
\end{equation}
   The factor of $4\pi$ is related with the (standard) use of Gaussian units
for the polarizabilities but natural units in field-theoretical calculations.
   Clearly the electric polarizability is expected to be a positive quantity
and its magnitude provides a measure of the ``stiffness'' or resistance to
deformation of the system.

   Similarly, when a static and uniform magnetizing field $\mathbf{H}$ is
applied to such an elementary system there are two competing mechanisms at
work.
   On the one hand, if the fundamental constituents of the system themselves
possess intrinsic magnetic dipole moments, they tend to align in the
direction of the applied field, producing a positive (paramagnetic) effect.
   On the other hand, by Lenz' law an applied field induces currents which
produce an induced magnetic moment opposite to this field, yielding a
negative (diamagnetic) effect.
   Using the definition
\begin{equation}
\mathbf{m}=4\pi\beta_M\mathbf{H},
\end{equation}
corresponding to a potential energy
\begin{equation}
U_H=-{1\over 2}4\pi\beta_M\mathbf{H}^2,
\quad\textnormal{\it i.e.},\quad
\mathbf{m}=-{\partial U_H\over \partial \mathbf{H}},
\end{equation}
the magnetic polarizability can be either positive of negative
depending on which mechanism is dominant.

   An equivalent means of expressing these results is in terms of the energy
density of a dilute macroscopic collection of such polarizable particles
randomly distributed within a volume $V$.
   If $N(\mathbf{x})$ is the number density of these particles then the energy
density of the (dilute) system including polarizability effects is given by
\begin{equation}
u(\mathbf{x})={1\over 2}\mathbf{E}_0^2(\mathbf{x})(1-4\pi N(\mathbf{x})\alpha_E)
+{1\over 2}\mathbf{H}_0^2(\mathbf{x})(1-4\pi N(\mathbf{x})\beta_M),
\end{equation}
where $\mathbf{E}_0$ and $\mathbf{H}_0$ denote given external electric and
magnetic fields applied to the medium.
   For the moment we assume a constant and uniform number density $N$.
   Comparing with the definition of the dielectric constant $\epsilon$ and
magnetic susceptibility $\mu$,
\begin{equation}
u={1\over 2}\epsilon\mathbf{E}^2+{1\over 2}\mu\mathbf{H}^2,
\end{equation}
where
\begin{equation}
\mathbf{E}=(1-4\pi N\alpha_E)\mathbf{E}_0\quad{\rm and}\quad\mathbf{H}
=(1-4\pi N\beta_M)\mathbf{H}_0
\end{equation}
are the net fields in the presence of the induced dipoles, we see that at the
classical level we identify at leading order in the number density
\begin{equation}
\epsilon=1+4\pi N\alpha_E\quad{\rm and}\quad \mu=1+4\pi N\beta_M.
\label{eq:pl}
\end{equation}
   Thus Eq.\ (\ref{eq:pl}) provides a way to determine the polarizabilities via
measurement of $\epsilon,\mu$ or of the index of refraction, which is given by
\begin{equation}
n=\sqrt{\epsilon\mu}\simeq 1+2\pi N(\alpha_E+\beta_M).
\end{equation}
   Since $N\alpha_E,N\beta_M$ must be dimensionless, we see that the electric
and magnetic polarizabilities $\alpha_E,\,\beta_M$ have units of volume.

   A simple example of this phenomenon is provided by the hydrogen atom.
   In a classical picture, if we represent the atom as an electron bound to a
proton via a harmonic oscillator potential with frequency $\omega_0$,
application of an electric field $\mathbf{E}$ generates a charge separation
\begin{equation}
\boldsymbol{\delta}=\mathbf{r}_e-\mathbf{r}_p
=-{e\,\mathbf{E}\over m_r\omega_0^2},
\end{equation}
where $e$ is the proton charge and $m_r=m_pm_e/(m_p+m_e)$ is the reduced mass.
   The resultant electric dipole moment is
\begin{equation}
\mathbf{p}=-e\,\boldsymbol{\delta}={e^2\mathbf{E}\over m_r\omega_0^2}
\end{equation}
which, using
\begin{equation}
U_E=-\frac{1}{2}\mathbf{p}\cdot\mathbf{E}
\end{equation}
for the potential energy of an induced dipole moment in an external field,
corresponds to an electric polarizability
\begin{equation}
\alpha_E={\alpha_{\rm em}\over m_r\omega_0^2},
\end{equation}
where $\alpha_{\rm em}=e^2/4\pi\approx 1/137$ is the fine-structure constant.
   Utilizing a value for $\omega_0$ corresponding to the rotational frequency
of the hydrogen ground state---$\omega_0=\alpha_{\rm em}^2m_r/2$---we have then
\begin{equation}
\alpha_E^H={4\over \alpha_{\rm em}^3m_r^3}={3\over \pi}V_H,
\label{alphaEHclassical}
\end{equation}
where $V_H=4\pi a_0^3/3$, with $a_0=1/(m_r\alpha_{\rm em})$ being the Bohr radius,
is the volume of the hydrogen atom.
   In a fully quantum mechanical calculation, the hydrogen atom polarizability
can be exactly calculated.
   Using the interaction potential
\begin{equation}
V_\textnormal{int}=-(-e\,\mathbf{r})\cdot\mathbf{E}=e\,\mathbf{r}\cdot\mathbf{E},
\end{equation}
we have in second-order perturbation theory
\begin{equation}
\Delta E_0^{(2)}=-e^2\sum_{n\neq 0}{\langle 0|\mathbf{r}\cdot\mathbf{E}|n\rangle
\langle n|\mathbf{r}\cdot\mathbf{E}|0\rangle\over E_n-E_0}.
\end{equation}
   Equating $\Delta E_0^{(2)}=U_E=-2\pi\alpha_E^H\mathbf{E}^2$, choosing
$\mathbf{E}$ along the $z$-axis, and performing the intermediate state summation,
we find \cite{mzb}
\begin{equation}
\alpha_E^H=2\alpha_{\rm em}\sum_{n\neq 0}{\langle 0|z|n\rangle\langle n|z|0\rangle\over E_n-E_0}
=\frac{9}{2 a_0^3}={27\over 8\pi}V_H,
\end{equation}
so that again we determine that $\alpha_E^H\sim V_H$.

   The magnetic polarizability of the hydrogen atom is much smaller than its
electric counterpart.
   Classically $\beta_M^H$ can be estimated via turning on a uniform magnetizing field
$\mathbf H=H (t)\,\mathbf{\hat{e}}_z$ in the vicinity of the atom.
   If the center of the atom is at the origin and one considers a closed
circle of radius $\rho$ in the $(x,y)$ plane concentric with the origin,
the changing magnetic field leads via Faraday's law to a circulating electric
field $\mathbf{E}=E(\rho,t)\,\mathbf{\hat{e}_\phi}$ \cite{fnm},
\begin{equation}
2\pi\rho E(\rho,t)=-\frac{d}{dt}\left[\pi \rho^2H(t)\right]\quad
\textnormal{yielding}\quad E=-{\rho\over 2}{dH\over dt}.
\end{equation}
   The resulting torque on an electron generates an angular momentum change
\begin{equation}
{d\mathbf{L}\over dt}
=\rho\, \mathbf{\hat{e}_\rho}\times (-e E\, \mathbf{\hat{e}_\phi})
=e{\rho^2\over 2}{dH\over dt}\,\mathbf{\hat{e}}_z.
\end{equation}
   Integrating with respect to time from zero field, results in an extra angular momentum
\begin{equation}
\Delta\mathbf{L}=e \frac{\rho^2}{2}H\,\mathbf{\hat{e}}_z,
\end{equation}
producing an additional orbital magnetic moment
\begin{equation}
\Delta\boldsymbol{\mu}=-\frac{e}{2m_r}\Delta\mathbf{L}
=-{e^2\langle r^2\rangle \over 6m_r}H\,\mathbf{\hat{e}}_z,
\end{equation}
where we have used spherical symmetry to write
$\langle x^2+y^2\rangle={2\over 3}\langle r^2\rangle$.
   According to Lenz' law, the added moment is indeed opposite to the
magnetic field.
   Using $U_M=-\Delta\boldsymbol{\mu}\cdot\mathbf{H}/2$, we
find then a negative magnetic polarizability
\begin{equation}
\beta_M^H=-{\alpha_{\rm em}a_0^2\over 6m_r}
=-{1\over 6\alpha_{\rm em}m_r^3}=-{1\over 24}\alpha_{\rm em}^2\alpha_E^H\sim 10^{-6}\,V_H,
\label{betaMHclassical}
\end{equation}
with $\alpha_E^H$ of Eq.~(\ref{alphaEHclassical}).
   The analogous result can be obtained quantum mechanically by considering
a static and uniform external magnetizing field in the positive $z$ direction,
$\mathbf H= H\mathbf{\hat{e}}_z$, giving rise to a perturbation
\begin{equation}
V_\textnormal{int}=\frac{e}{2m_r}l_z H+\frac{e^2}{8m_r}(x^2+y^2)H^2=H_1+H_2,
\end{equation}
where $l_z$ is the $z$ component of the relative orbital angular momentum operator.
   Applying second-order perturbation theory results in
\begin{equation}
\Delta E_0^{(2)}=\langle 0|H_2|0\rangle-
\sum_{n\neq 0}{|\langle n|H_1|0\rangle|^2 \over E_n-E_0}
=\frac{2}{3}\langle r^2\rangle
\frac{e^2 H^2}{8m_r},
\end{equation}
since $l_z|0\rangle=0$.
   Equating $\Delta E_0^{(2)}=U_M=-2\pi\beta_M^H\mathbf{H}^2$ and comparing
with Eq.~(\ref{betaMHclassical}), we find exactly the same result
for $\beta_M^H$ as found in the classical physics derivation.

\subsection{Hadron Polarizabilities}
\begin{figure}
\begin{center}
\epsfig{file=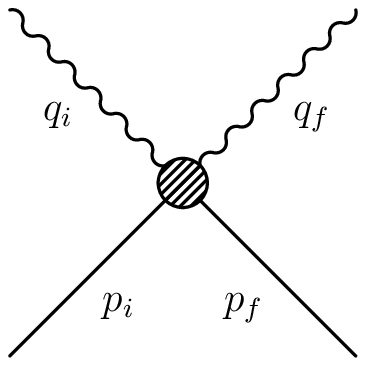,width=0.3\textwidth}
\end{center}
\caption[test]{\label{figurekin} Compton scattering kinematics}
\end{figure}

   Next we move to consider the polarizabilities of strongly interacting
particles, namely, hadrons.
   We begin with a simple charged spinless particle such as, e.g., a
$\pi^+$.
   Because the particle is charged, the use of macroscopic samples and a
constant external electric field is not feasible.
   Instead one uses the fields generated in a Compton scattering process,
$\gamma(q_i)+\pi^+(p_i)\to\gamma(q_f)+\pi^+(p_f)$ (see Fig.~\ref{figurekin}).
   In this case when the wavelength of the photon is much larger than that of
the target, any structure of the particle cannot be resolved and it is characterized
only by its charge $e$ and mass $M$.
   The Hamiltonian is then the simple Schr\"odinger form relevant for a
charged particle in the presence of an external electromagnetic four-vector
potential $A^\mu=(\phi,\mathbf{A})$  \cite{jdj}
\begin{equation}
H_0={(\mathbf{p}-e\mathbf{A})^2\over 2M}+e\phi.
\end{equation}
   Using the Lorenz condition $\partial_\mu A^\mu=0$ and the Coulomb gauge
$\phi=A^0=0$, the corresponding transition amplitude has the Thomson form
\begin{equation}
T_{0}^\textnormal{Compton}
=-{e^2\over M}\,\mathbf{\hat{\boldsymbol\epsilon}}_f^*\cdot\mathbf{\hat{\boldsymbol\epsilon}}_i,
\end{equation}
leading to the well-known (laboratory frame) Thomson cross section \cite{fnm,brh}
\begin{equation}
{d\sigma_{\rm Compton}\over d\Omega}
={\alpha_{\rm em}^2\over 2M^2}{\omega_f^2\over \omega_i^2}(1+\cos^2\theta).
\end{equation}
   The total Thomson cross section, obtained by taking the limit $\omega_i\to 0$ and
integrating over the entire solid angle, reproduces the classical Thomson scattering
cross section denoted by $\sigma_T$,
\begin{equation}
\label{sigmat}
\sigma_T=\frac{8\pi}{3}\frac{\alpha^2_{\rm em}}{M^2}.
\end{equation}
   Numerical values of $\sigma_T$ for the electron, charged
pion, and the proton are shown in Table \ref{tcs}.

\begin{table}
\caption[test]{\label{tcs} Thomson cross section $\sigma_T$ for
the electron, charged pion, and proton}
\begin{center}
\renewcommand{\arraystretch}{1.3}
\begin{tabular}{ll}
Particle&$\sigma_T$\\
\hline
\hline
Electron & 0.665 barn\\
Pion & 8.84 $\mu$barn\\
Proton & 197 nbarn\\
\hline
\hline
\end{tabular}
\end{center}
\end{table}

   As the wavelength becomes smaller and comparable to the size of the target,
the particle begins to be resolved and the simple Schr\"odinger Hamiltonian $H_0$
is augmented by a form which includes polarizability corrections \cite{Klein:1955zz},
\begin{equation}
H=H_0-{1\over 2}4\pi\alpha_E\mathbf{E}^2-{1\over 2}4\pi\beta_M\mathbf{H}^2+\ldots.
\label{eq:gv}
\end{equation}
The Compton amplitude becomes then
\begin{equation}
T_{\rm Compton}=\mathbf{\hat{\boldsymbol\epsilon}}_f^*\cdot\mathbf{\hat{\boldsymbol\epsilon}}_i
\left(-\frac{e^2}{M}+4\pi\alpha_E\omega_f\omega_i\right)
+4 \pi\beta_M\mathbf{\hat{\boldsymbol\epsilon}}_f^*\times\mathbf{q}_f\cdot\,
\mathbf{\hat{\boldsymbol\epsilon}}_i\times\mathbf{q}_i+\ldots,\label{eq:bv}
\end{equation}
while the cross section is
\begin{align}
{d\sigma_{\rm Compton}\over d\Omega}&=
{1\over M^2}{\omega_f^2\over \omega_i^2}\left\{{\alpha_{\rm em}^2\over 2}(1+\cos^2\theta)
-\alpha_{\rm em}M\omega_f\omega_i\left[{1\over 2}(\alpha_E+\beta_M)(1+\cos\theta)^2
\right.\right.\nonumber\\
&\quad\left.\left.
+{1\over 2}(\alpha_E-\beta_M)(1-\cos\theta)^2\right]+\ldots\right\},
\label{eq:jn}
\end{align}
so that $\alpha_E,\beta_M$ can be determined by a careful measurement of the
angular distribution in Compton scattering.

\subsection{Proton Polarizabilities}
   In the case of the proton, there exists an additional contribution due to
the feature that the proton possesses a magnetic moment (mm).
   Thus a piece
\begin{equation}
H_\textnormal{mm}=-\boldsymbol{\mu}_p\cdot\mathbf{H}
\end{equation}
must be appended to the interaction Hamiltonian.
   The magnetic moment is given by
\begin{equation}
\boldsymbol{\mu}_p=g_p\, \mu_N \mathbf{S},
\end{equation}
where $g_p=5.59$ is the proton gyromagnetic ratio,
$\mu_N=e/(2m_p)$ the nuclear magneton, and
$\mathbf{S}$  the spin operator.
   The corresponding relativistic cross section in the absence of the
polarizability terms is the Powell cross section \cite{pow} and an experimental
search for polarizabilities seeks deviations from this form.
   An important consideration here is the photon energy.
   The size of such polarizability terms compared to the leading Thomson amplitude is
\begin{equation}
\sim 2 \alpha_E^p {m_p\omega^2\over \alpha_{\rm em}}\approx 33{\omega^2\over m^2_p},
\end{equation}
where we made use of the empirical value $\alpha_E^p=11.2\times 10^{-4}\,{\rm fm}^3$.
   In order to produce a sizable polarizability effect without bringing in large contributions
from higher-order terms the photon energy is ideally in the 50--100 MeV range.
   However, in this case and even more so for the case of larger energies
it is necessary to have an estimate for the contribution of higher-order effects and this
is generally provided by various dispersion relation codes (see, e.g., Refs.~\cite{lvo,pas})
or effective-field-theory calculations
(see, e.g., Refs.\ \cite{Beane:2002wn,Lensky:2009uv,McGovern:2012ew}).
   In this fashion a series of measurements on the proton
\cite{Federspiel:1991yd,Zieger:1992jq,Mac,Bla,Olm}
has yielded the rather precise numbers \cite{pdg}:
\begin{equation}
\alpha_E^p=(11.2\pm 0.4)\times 10^{-4}\,{\rm fm}^3\quad{\rm and}\quad
\beta_M^p=(2.5\mp 0.4)\times 10^{-4}\,{\rm fm}^3.
\label{eq:wr}
\end{equation}
   In terms of the proton volume---$V_p={4\pi\over 3}\langle (r_E^p)^2\rangle^{3\over 2}$---we have then
$\alpha_E^p\sim 4\times 10^{-4}\,V_p$, showing that the proton is much stiffer and more strongly bound
than the hydrogen atom.

   Another interesting feature of these experimental results comes from the magnetic polarizability.
   The $\Delta$ pole diagram makes a rather strong para\-magnetic contribution \cite{Mukhopadhyay:1993zx}
\begin{equation}
\beta_M^p(\Delta-{\rm pole})\sim 10\times 10^{-4}\,{\rm fm}^3
\end{equation}
so that the fact that the experimental number is only about 25 \% of this value reveals a
very strong diamagnetic contribution, presumably from the meson cloud.
   Such a picture is qualitatively supported by the Skyrme model \cite{Sch}.

\subsection{Dispersion Relations}
   An alternative approach to the size of the polarizabilities comes from the use of
causality via dispersion relations \cite{GellMann:1954db}.
   In the case of forward Compton scattering, we can write
\begin{equation}
T_{\rm forward}=4\pi f_0(\omega)\hat{\boldsymbol\epsilon}_f^*\cdot\hat{\boldsymbol\epsilon}_i
+4\pi ig_0(\omega)\boldsymbol{\sigma}\cdot\hat{\boldsymbol\epsilon}_f^*\times\hat{\boldsymbol\epsilon}_i,
\label{eq:bh}
\end{equation}
where the subscripts 0 refer to $\theta=0^\circ$.
   Equation (\ref{eq:bh}) needs to be evaluated between Pauli spinors of the initial
and final nucleons, respectively.
   In the case of the spin-averaged (scalar) amplitude $f_0(\omega)$, the imaginary part is
given by the optical theorem via
\begin{equation}
{\rm Im}\left[f_0(\omega)\right]={\omega\over 4\pi}\sigma_{\rm tot}(\omega),
\end{equation}
where $\sigma_{\rm tot}(\omega)$ is the total photoabsorption cross section.
   Using $f_0^\ast(\omega)=f_0(-\omega)$ and applying Titchmarsh's theorem \cite{Titchmarsh}
to $(f_0(\omega)-f_0(0))/\omega^2$, the corresponding real part of the proton scattering
amplitude can be written as
\begin{equation}
{\rm Re}\left[f_0^p(\omega)\right]
=-{\alpha_{\rm em}\over m_p}+{\omega^2\over 2\pi^2}P\int_0^\infty
d\omega'{\sigma_{\rm tot}^p(\omega')\over {\omega'}^2-\omega^2},
\end{equation}
where we have performed a subtraction since the presence of the Thomson amplitude indicates
that an unsubtracted dispersion relation does not converge.
   Expanding the scattering amplitude below the pion production threshold via
\begin{equation}
f^p_0(\omega)=-{\alpha_{\rm em}\over m_p}+\left(\alpha_E^p+\beta_M^p\right)\omega^2+\ldots
\end{equation}
leads to the Baldin sum rule \cite{bsr}
\begin{equation}
\alpha_E^p+\beta_M^p={1\over 2\pi^2}\int_0^\infty d\omega{\sigma_{\rm tot}^p(\omega)\over {\omega}^2}
\end{equation}
which relates the sum of electric and magnetic polarizabilities to a weighted integral over
the total photoabsorption cross section.
   In the case of the proton, using experimental cross section numbers, rather
precise values have been calculated:
\begin{equation}
\alpha_E^p+\beta_M^p=\left\{\begin{array}{l}
(13.69\pm 0.14)\times 10^{-4}\,{\rm fm}^3\,\cite{Babusci:1997ij},\\
(14.0\pm 0.5)\times 10^{-4}\,{\rm fm}^3\,\cite{Levchuk:1999zy},\\
(13.8\pm 0.4)\times 10^{-4}\,{\rm fm}^3\,\cite{Olm}.
\end{array}
\right.
\label{eq:bsrp}
\end{equation}
   In order to separate the electric and magnetic polarizabilities, one can use a
backward dispersion relation, the Bernabeu-Ericson-Ferro Fontan-Tarrach (BEFT) sum rule,
to determine $\alpha_E-\beta_M$, which takes the form \cite{Bernabeu:1974zu,Bernabeu:1977hp}
\begin{equation}
(\alpha_E^p-\beta_M^p)^{\rm BEFT}=(\alpha_E^p-\beta_M^p)_s+(\alpha_E^p-\beta_M^p)_t.
\end{equation}
   Here, the $s$-channel contribution is given by
\begin{equation}
(\alpha_E^p-\beta_M^p)_s=
{1\over 2\pi^2}\int_0^\infty {d\omega\over \omega^2}\left(1+{2\omega\over m_p}\right)^{1\over 2}
\big(\sigma^p(\omega,{\rm yes})-\sigma^p(\omega,{\rm no})\big),
\label{eq:nh}
\end{equation}
where $\sigma(\omega,\Delta P)$ represents components of the photoabsorption cross section associated with
multipoles which change ($\Delta P={\rm yes}$), do not change ($\Delta P={\rm no}$) parity.
   Numerical evaluation of Eq.~(\ref{eq:nh}) is reasonably robust and yields \cite{Schumacher:2013hu}
\begin{equation}
(\alpha_E^p-\beta_M^p)_s=-5.0\times 10^{-4}\,{\rm fm}^3.
\end{equation}
   Calculation of the $t$-channel contribution requires summation over a complete set
of intermediate-state contributions to $p\bar{p}\rightarrow\,\mbox{hadrons}\,\rightarrow \gamma\gamma$.
   The lightest such state is $\pi\pi$ and is dominated by low-partial-wave contributions.
   Including only $S$- and $D$-wave pieces, we have \cite{Holstein:1994tw}
\begin{align}
(\alpha_E^p-\beta_M^p)_t&=
{1\over 16\pi^2}\int_{4M_\pi^2}^\infty {dt\over t^2}{16\over 4m_p^2-t}\sqrt{t-4M_\pi^2\over t}\nonumber\\
&\quad\times\left[f_+^{p\,0}(t)F_0^{0*}(t)
-\left(m_p^2-{t\over 4}\right)\left({t\over 4}-M_\pi^2\right)f_+^{p\,2}(t)F_0^{2*}(t)+\ldots\right],
\end{align}
where $f_+^{p\,J}(t)$ are partial-wave amplitudes for $p\bar{p}\rightarrow\pi\pi$
and $F_0^{J*}(t)$ are the corresponding amplitudes for $\pi\pi\rightarrow\gamma\gamma$.
   Past evaluations have yielded a range of values for this contribution, namely, from
$10.3\times 10^{-4}\,{\rm fm}^3$ to $16.1\times 10^{-4}\,{\rm fm}^3$ for the $S$-wave piece
and $-1.3 \times 10^{-4}\,{\rm fm}^3$ for the $D$ wave \cite{Holstein:1994tw}.
   In the case that the $S$ wave dominates, one approach is to use the $\sigma$-pole contribution
to estimate this value---
\begin{equation}
(\alpha_E^p-\beta_M^p)_t\simeq {g_{\sigma NN}{\cal M}(\sigma\rightarrow\gamma\gamma)\over 2\pi M_\sigma^2}.
\end{equation}
   Using a simple Nambu-Jona-Lasinio model, we find for the $\sigma NN$ coupling constant
$g_{\sigma NN}=g_{\pi NN}=13.12$ \cite{Baru:2010xn} and \cite{Schumacher:2011gi}
\begin{equation}
M_\sigma=\left({16\pi^2\over N_c}F_\pi^2+M_\pi^2\right)^{1\over 2},
\end{equation}
where $F_\pi=92.2$ MeV \cite{pdg} is the pion-decay constant and $N_c=3$ the number of colors.
Thus
\begin{equation}
{\cal M}(\sigma\rightarrow\gamma\gamma)={3\alpha_{\rm em}\over \pi F_\pi}
\left[\left({2\over 3}\right)^2+\left(-{1\over 3}\right)^2\right]
\end{equation}
and
\begin{equation}
(\alpha_E^p-\beta_M^p)_t=15.2 \times 10^{-4}\,{\rm fm}^3.
\end{equation}
   In combination with $(\alpha_E^p+\beta_M^p)=14.0\times 10^{-4}\,{\rm fm}^3$ \cite{Levchuk:1999zy},
this result yields
\begin{equation}
\alpha_E^p=12.1\times 10^{-4}\,{\rm fm}^3\quad{\rm and}\quad\beta_M^p=1.9\times 10^{-4}\,{\rm fm}^3
\end{equation}
in good agreement with direct measurements.
   However, a definitive numerical approach to the $t$-channel contribution remains to be achieved.

\subsection{Theoretical Approaches}
   That the size of the nucleon polarizabilities are reasonable can be seen from a simple
nonrelativistic harmonic oscillator valence quark model calculation \cite{Drechsel:1984nk}.
   Denoting the proton ground state by $|0\rangle_p$, we can use the result
\cite{Petrunkin:1964,Ericson:1973,Friar:1975}
\begin{align}
\alpha_E^p&={\alpha_{\rm em}\over 3m_p}\,{}_p\langle 0|r^2_E|0\rangle_p
+2\alpha_{\rm em}\sum_{n\neq 0}{|\langle n|D_z|0\rangle_p|^2\over E_n-E_0}
\end{align}
with the mean-square charge radius and electric dipole operators, respectively,
\begin{displaymath}
r^2_E=\sum_{i=1}^3q_i(\mathbf{r}_i-\mathbf{R})^2,\quad
\mathbf D=\sum_{i=1}^3q_i(\mathbf{r}_i-\mathbf{R}),
\end{displaymath}
where $\mathbf R$ is the center-of-mass coordinate.
   Here the oscillator frequency $\omega_0$ can be found in terms of the charge radius via
\begin{equation}
\omega_0=\frac{3}{m_p\langle (r_E^p)^2\rangle}\simeq 160 \,{\rm MeV}
\end{equation}
and the predicted electric polarizability becomes \cite{Holstein:1992xr}
\begin{equation}
\alpha_E^p={2\alpha_{\rm em} m_p\over 9}\langle(r_E^p)^2 \rangle^2\simeq 46\times 10^{-4}\,{\rm fm}^3
\end{equation}
which is in the ballpark but is about a factor of four too large.
   This overprediction can be understood in terms of the feature that the
predicted oscillator frequency, which measures the excitation energy, is somewhat too small.
   If a more reasonable value of $\sim$ 300 MeV is used, the number comes out about right.
   In any case we see that the size of the electric
polarizability---$\alpha_E\sim 10^{-3}\,V$---is not unreasonable.

\begin{figure}
\begin{center}
\epsfig{file=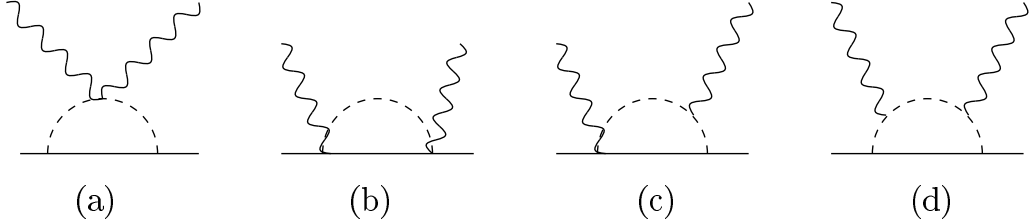,width=\textwidth}
\end{center}
\caption{\label{figure:diagrams} One-loop diagrams in HBChPT. Crossed diagrams are not shown.}
\end{figure}

\begin{figure}[htbp]
    \centering
        \includegraphics[width=\textwidth]{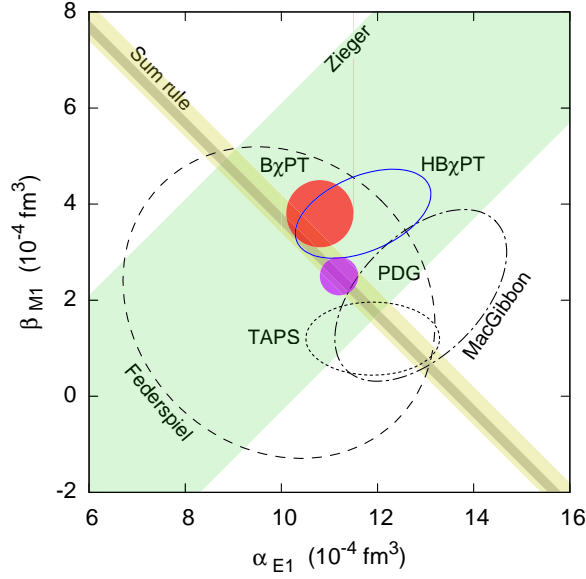}
    \caption{Electric and magnetic polarizabilities $\alpha_{E1}(=\alpha_E^p)$ and
$\beta_{M1}(=\beta_M^p)$
of the proton including an updated PDG value \cite{Krupina:2013dya}.
   The magenta blob represents the PDG summary of Eq.~(\ref{eq:wr}) \cite{pdg}.
   The red blob represents the RChPT calculation of Ref.\ \cite{Lensky:2009uv} and the
blue ellipse the HBChPT calculation of Ref.\ \cite{McGovern:2012ew}, respectively.
   Sum rule indicates the Baldin sum rule evaluations of Eq.~(\ref{eq:bsrp}) from
Refs.~\cite{Olm} (broad band) and \cite{Babusci:1997ij} (narrow band), respectively.
   The experimental results are from Federspiel et al.~\cite{Federspiel:1991yd},
Zieger et al.~\cite{Zieger:1992jq}, MacGibbon et al.~\cite{Mac}, and TAPS \cite{Olm}.
\label{fig:alphabetav2}
}
\end{figure}

   The treatment of the magnetic polarizability is more challenging, but in any case
a constituent quark model is not expected to be successful, as long
as it does not deal with the pion cloud, which should be an important component of
any realistic nucleon model.
   In this regard the use of chiral perturbation theory (ChPT)
\cite{Weinberg:1978kz,Gasser:1983yg,Gasser:1987rb} is promising
(see, e.g., Refs.~\cite{Ecker:1994gg,Bernard:1995dp,Scherer:2002tk,Scherer:2012xha}
for an introduction).
   In ChPT physical observables are calculated perturbatively in terms of
a momentum expansion in $q/\Lambda_\chi$.
   The chiral-symmetry-breaking scale $\Lambda_\chi$ is $4\pi F_\pi \sim$ 1 GeV
\cite{Manohar:1983md,Donoghue:1984gg} and $q$ collectively stands for a quantity small
in comparison to $\Lambda_\chi$
such as the pion mass, small external four-momenta of the pion,
and small external three-momenta of the nucleon.
   The first such calculation of the proton polarizability was that of Bernard, Kaiser,
and Mei{\ss}ner, which was performed using relativistic ChPT \cite{Gasser:1987rb}
at the one-loop level.
   The results were obtained in a closed (Feynman parameter) integral form
and their chiral expansion in terms of $\mu=M_\pi/m_N$ reads \cite{Bernard:1991rq}
\begin{equation}
\begin{split}
\alpha_E^p&={\alpha_{\rm em}g_A^2\over 48\pi^2F_\pi^2 m_N}\left[{5\pi\over 2\mu}
+18\ln\mu+{33\over 2}+{\cal O}(\mu)\right]=7.9\times 10^{-4}\,{\rm fm}^3,\\
\beta_M^p&=
{\alpha_{\rm em}g_A^2\over 48\pi^2F_\pi^2m_N}\left[{\pi\over 4\mu}
+18\ln\mu+{63\over 2}+{\cal O}(\mu)\right]=-2.3\times 10^{-3}\,{\rm fm}^3,\\
\end{split}
\label{eq:vf}
\end{equation}
   where the numerical values refer to the full expressions.
   Note that the expressions of Eq.~(\ref{eq:vf}) for the polarizabilities
contain no free parameters, {\it i.e.}, they are entirely expressed in terms of $M_\pi$, $m_N$, $F_\pi$, and
the axial-vector coupling constant $g_A=1.27$.
   However, manifestly Lorentz-invariant (or relativistic) ChPT (RChPT) seemingly had
a problem concerning power counting when loops containing internal nucleon
lines come into play.
   In some cases, a diagram apparently has contributions which are of lower order than
determined by the power counting \cite{Gasser:1987rb}.
   Therefore, the calculation was redone using the methods of heavy-baryon ChPT (HBChPT)
\cite{Jenkins:1990jv,Bernard:1992qa}.
   In this approach, one evaluates the spin-averaged values of the four Compton scattering
diagrams shown in Fig.~\ref{figure:diagrams} in terms of the representation
\begin{equation}
T_{\rm Compton}^{\rm spin-averaged}
=\mathbf{\hat{\boldsymbol\epsilon}}_f^*\cdot\mathbf{\hat{\boldsymbol\epsilon}}_i\, A_1
+\mathbf{\hat{\boldsymbol\epsilon}}_f^*\cdot\mathbf{\hat q}_i\,
\mathbf{\hat{\boldsymbol\epsilon}}_i\cdot\mathbf{\hat q}_f\, A_2.
\end{equation}
   The results at ${\cal O}(q^3)$ can be expressed as a Born plus one-loop contribution \cite{Bernard:1992qa}
\begin{equation}
\begin{split}
A_{1,\,\rm Born}^p&=-{e^2\over m_N},\quad
A_{2,\,\rm Born}^p={e^2\omega\over m_N^2},\\
A_{1,\,\rm loop}^p&={e^2 g_A^2\omega^2\over 192\pi F_\pi^2M_\pi}(10+\cos\theta),\quad
A_{2,\,\rm loop}^p=-{e^2 g_A^2\omega^2\over 192\pi F_\pi^2M_\pi}.
\end{split}
\end{equation}
   Using
\begin{displaymath}
\mathbf{\hat{\boldsymbol\epsilon}}_f^*\times\mathbf{\hat{q}}_f\cdot\,
\mathbf{\hat{\boldsymbol\epsilon}}_i\times\mathbf{\hat{q}}_i
=\mathbf{\hat{\boldsymbol\epsilon}}_f^*\cdot\mathbf{\hat{\boldsymbol\epsilon}}_i\cos\theta
-\mathbf{\hat{\boldsymbol\epsilon}}_f^*\cdot\mathbf{\hat q}_i\,
\mathbf{\hat{\boldsymbol\epsilon}}_i\cdot\mathbf{\hat q}_f
\end{displaymath}
and comparing with the Compton amplitude in Eq.\ (\ref{eq:bv}), we reproduce the Thomson amplitude
and identify the polarizabilities at ${\cal O}(q^3)$ as
\begin{equation}
\begin{split}
\alpha_{E,{\rm HB},q^3}^p&={10\alpha_{\rm em}g_A^2\over 192\pi F_\pi^2 M_\pi}=12.6\times 10^{-4}\,{\rm fm}^3,\\
\beta_{M,{\rm HB},q^3}^p&={\alpha_{\rm em}g_A^2\over 192\pi F_\pi^2 M_\pi}=1.26\times 10^{-4}\,{\rm fm}^3
\end{split}
\end{equation}
which reproduces the leading $1/M_\pi$ term in the relativistic ChPT expression of Eq.~(\ref{eq:vf})
and agrees quite well with the experimental values of Eq.\ (\ref{eq:wr}).
   However, it seems clear that this agreement is somewhat accidental since an ${\cal O}(q^4)$ estimate
by Bernard, Kaiser, Schmidt, and Mei{\ss}ner yielded the modified values
\cite{Bernard:1993bg,Bernard:1993ry}
\begin{equation}
\alpha_{E,{\rm HB},q^4}^p=(10.5\pm 2.0)\times 10^{-4}\,{\rm fm}^3\quad{\rm and}
\quad\beta_{M,{\rm HB},q^4}^p=(3.5\pm 3.6)\times 10^{-4}\,{\rm fm}^3.
\label{eq:abhb4}
\end{equation}
   In particular, the results of Eq.~(\ref{eq:abhb4}) include an estimate of low-energy constants
(LECs) entering at ${\cal O}(q^4)$.
   In addition, the inclusion of the $\Delta(1232)$ resonance in either the ``small scale expansion''
($\epsilon$ expansion) \cite{Hemmert:1996rw} or the $\delta$ expansion
\cite{McGovern:2012ew,Pascalutsa:2002pi} has been shown to make significant modifications
to the lowest-order results.
   In the meantime, the power-counting problem in RChPT has been solved in the framework of
{\it infrared regularization} (IR) \cite{Becher:1999he,Schindler:2003xv} and the extended on-mass-shell
(EOMS) scheme \cite{Gegelia:1999gf,Fuchs:2003qc}.
   Covariant calculations have been performed at next-to-next-to-leading order including the
$\Delta(1232)$ resonance \cite{Lensky:2009uv} and also in the framework of the linear sigma
model \cite{Metz:1996fn}.
   In particular, it was stressed in Ref.\ \cite{Lensky:2009uv} that the leading-order behavior of the
forward-scattering combination ($\alpha_E^p+\beta_M^p$) essentially reflects the Kroll-Ruderman theorem of
pion photoproduction \cite{Kroll:1953vq}, whereas the backward-scattering combination
($\alpha_E^p-\beta_M^p$)
relies on chiral symmetry.
   Figure \ref{fig:alphabetav2} summarizes the present experimental and theoretical
situation regarding the scalar polarizabilities of the proton \cite{Krupina:2013dya}.

\subsection{Neutron Polarizability}

   In chiral perturbation theory, the electric and magnetic polarizabilities
are expected to be primarily isoscalar quantities, so that the proton and neutron values should
be similar.
   However, on the experimental side, the absence of a neutron target means that the measurement
of the neutron polarizabilities is a challenging prospect.
   Nevertheless, there exist results from a number of approaches.

   One technique is to extract the electric polarizability from slow-neutron
electromagnetic scattering in the Coulomb field of heavy nuclei,
for which the cross section is \cite{Lvov:1993fp}
\begin{equation}
\frac{d\sigma_{\rm em\, pol}}{d\Omega}
=4\pi^2Z^2\alpha_{\rm em}m_np_n{\rm Re}(a)\left[\alpha_E^n\sin\frac{\theta}{2}
-\frac{2\pi \alpha_{\rm em}\kappa_n^2}{m_n^3}\left(1-\sin\frac{\theta}{2}\right)\right],
\end{equation}
where $p_n$ is the absolute value of the neutron three-momentum, $-a$ is the amplitude for hadronic
scattering by the nucleus,
$Z$ is the nuclear charge, and $\kappa_n=-1.913$ is the (anomalous) neutron magnetic moment.
   In this way an experiment at Oak Ridge National Laboratory has yielded a result
\cite{Schmiedmayer:1991zz}
\begin{equation}
\label{alphaEnnscattering}
\alpha_E^n=(12.0\pm 1.5\pm 2.0)\times 10^{-4}\,{\rm fm}^3,
\end{equation}
in good agreement with the proton value---Eq.\ (\ref{eq:wr}).
   An alternative method to extract the neutron polarizabilities is to utilize
quasi-free Compton scattering on the neutron bound in the deuteron, $\gamma d\to \gamma np$,
which has given
\begin{equation}
\alpha_E^n=\left\{\begin{array}{ll}(10.7^{+3.3}_{-10.7})\times 10^{-4}\,{\rm fm}^3\,&\cite{Rose:1990zz},\\
(12.5\pm 2.3)\times 10^{-4}\,{\rm fm}^3&\cite{Kossert:2002ws},\end{array}\right.
\end{equation}
in good agreement with Eq.~(\ref{alphaEnnscattering}) determined via neutron scattering.
   This value concurs also with that extracted by the use of coherent Compton scattering from the deuteron,
$\alpha_E^n=(8.8\pm2.4\pm3.0) \times 10^{-4}\,{\rm fm}^3$ \cite{Lundin:2002jy}.
   The PDG average of the above results is \cite{pdg}
\begin{equation}
\alpha_E^n=(11.6\pm1.5) \times 10^{-4}\,{\rm fm}^3.
\end{equation}
   The Baldin sum rule \cite{bsr} for the neutron has been evaluated to be
\cite{Levchuk:1999zy}
\begin{equation}
\alpha_E^n+\beta_M^n
={1\over 2\pi^2}\int_0^\infty d\omega{\sigma_{\rm tot}^n(\omega)\over {\omega}^2}
=(15.2\pm 0.5)\times 10^{-4}\,{\rm fm}^3.
\end{equation}
   Using
\begin{equation}
\beta_M^n=\left\{\begin{array}{ll}(2.7\pm 1.8^{+1.3}_{-1.6})\times 10^{-4}\,{\rm fm}^3&\cite{Kossert:2002ws},\\
(6.5\pm2.4\pm 3.0)\times 10^{-4}\,{\rm fm}^3&\cite{Lundin:2002jy},
\end{array}\right.
\end{equation}
where both determinations make use of the Baldin sum rule, the PDG average is given by \cite{pdg}
\begin{equation}
\beta_M^n=(3.7\pm 2.0) \times 10^{-4}\,{\rm fm}^3,
\end{equation}
which is by about 50 \% larger than the proton value---Eq.\ (\ref{eq:wr})---but also
has a larger error bar.
   We conclude that the electromagnetic polarizabilities $\alpha_E,\beta_M$
are predominantly of isoscalar nature, and that
the relative isovector contribution appears to be larger for the magnetic polarizability
than for the electric polarizability.
   For a further discussion of how to extract the neutron polarizabilities see, e.g., Ref.\
\cite{Phillips:2009af} and references therein.
   Theoretical values have also been predicted for the polarizabilities of the hyperons,
 but it is unlikely that they will soon be measured \cite{Bernard:1992xi}.

\subsection{Charged-Pion Polarizability}

   A good deal of work has also been done involving the charged-pion polarizabilities.
   In this case, a valence quark picture is not expected to work well since it does not
capture the Goldstone nature of the pion.
   Instead a chiral perturbative technique, wherein the Goldstone nature of the pion
{\it is} exhibited, should be expected to be a reasonable approach and this turns
out to be the case \cite{Gasser:1983yg,Gasser:1984gg}.

   We define the second-rank Compton tensor as
\begin{equation}
T^{\mu\nu}(p_f,q_f;p_i,q_i)=i\int d^4x\, e^{-iq_i\cdot x}\langle\pi^+(p_f)|
T[J_{\rm em}^\mu (x) J_{\rm em}^\nu(0)]|\pi^+(p_i)\rangle,
\end{equation}
where $T$ denotes the covariant time-ordered product and $J_{\rm em}$ is
the electromagnetic current operator in units of the elementary charge.
   Because of translation invariance, four-momentum conservation is implied,
$p_f+q_f=p_i+q_i$.
   For real Compton scattering (RCS), $q_i^2=q_f^2=0$, the one-loop charged-pion
tensor is given by \cite{Pervushin:1974kc,Bijnens:1987dc}
\begin{align}
T^{\mu\nu}_{\rm RCS}&=-{\left( 2p_f + q_f\right)^\nu \left( 2 p_i + q_i\right)^\mu \over
\left( p_i+q_i\right)^2 - M^2_\pi} - {\left( 2p_f - q_i\right)^\mu \left( 2p_i -
q_f\right)^\nu\over \left( p_i - q_f\right)^2 - M^2_\pi} + 2g^{\mu\nu} \nonumber\\
&\quad+ \zeta \left( q_f^\mu q_i^\nu - g^{\mu\nu} q_i \cdot q_f\right) + \ldots,
\end{align}
where the first line includes the Born terms while the second line includes the
structure dependence.
   Comparing $T_{\rm RCS}=e^2\epsilon_{i\mu}\epsilon_{f\nu}^\ast T^{\mu\nu}_{\rm RCS}$ with the definitions for
electric, magnetic polarizabilities---Eq.~(\ref{eq:bv})---and noting that
the normalization of the covariant expression differs by a factor of $2M_\pi$, we identify
\begin{equation}
\alpha_E^{\pi^+}=-\beta_M^{\pi^+}= {\alpha_{\rm em}\over 2M_\pi}\zeta.
\end{equation}
   That is, at the one-loop level the electric and magnetic polarizabilities are equal and opposite.
   As for the absolute size, we can relate the polarizability to the axial structure constant
$h_A$ in radiative charged-pion beta decay---$\pi^+\rightarrow e^+\nu_e\gamma$---defined via
\cite{Terentev:1972ix,Donoghue:1989sq}
\begin{align}
A_{\mu\nu} (p,q) &= \int d^4x\,e^{iq\cdot x} \langle 0 \left|T\left( J^{\rm
em}_\mu (x) J^{{\rm wk}}_{\nu,1-i2} (0)\right)\right| \pi^+ (p) \rangle \nonumber\\
&= - \sqrt{2}\,F_\pi {(p-q)_\nu\over (p-q)^2 - M^2_\pi}
\langle\pi^+(p-q)\left| J^{\rm em}_\mu\right|\pi^+(p)\rangle
+ \sqrt{2}\,F_\pi g_{\mu\nu} \nonumber\\
&\quad- h_A \left[(p-q)_\mu q_\nu - g_{\mu\nu} q \cdot (p-q)\right]
- r_A (q_\mu q_\nu - g_{\mu\nu} q^2) \nonumber\\
&\quad + ih_V\epsilon_{\mu\nu\alpha\beta} q^\alpha p^\beta.
\label{eq:ku}
\end{align}
   In Eq.\ (\ref{eq:ku}), the second line is associated with the Born diagram
together with a term required for gauge invariance, while the structure-dependent pieces,
involving $h_A,r_A,h_V$ appear in the third and fourth lines, with the subscript $V,A$ indicating
its connection with the weak vector, axial-vector currents, respectively.
   (Note that the form factor $r_A$ vanishes for radiative decay to a real photon and
only comes into play for the Dalitz decay mode $\pi^+\rightarrow e^+\nu_ee^+e^-$.)
   Chiral symmetry relates $\zeta$ and $h_A$ via \cite{Donoghue:1993kw}
\begin{align}
h_V &=\left.\frac{N_c}{12\sqrt{2}\,\pi^2 F_\pi} \right|_{N_c=3},\nonumber\\
\frac{h_A}{h_V}&=32\pi^2\left(L^{r}_9(\mu) +L^{r}_{10}(\mu)\right),\nonumber\\
\frac{r_A}{h_V} &= 32\pi^2\left[L^{r}_9(\mu) - \frac{1}{192 \pi^2}
\left( \ln \frac{M^2_\pi}{\mu^2} + 1 \right)\right],\nonumber\\
\alpha_E^{\pi^+} &= \frac{\alpha_{\rm em}}{2M_\pi}\zeta
= \frac{\alpha_{\rm em}}{8\pi^2M_\pi F^2_\pi}\frac{h_A}{h_V}.
\label{eq:alphapiplus}
\end{align}
   Here, $h_V$ arises from the anomaly and is exactly predicted at ${\cal O}(q^4)$
\cite{Wess:1971yu,Witten:1983tw,Bijnens:1993xi}.
   The ratio $h_A/h_V$ is given in terms of a linear combination of LECs of the
${\cal O}(q^4)$ Lagrangian \cite{Gasser:1984gg}.
   The renormalization scale is denoted by $\mu$, but the linear combination
$L^{r}_9(\mu) +L^{r}_{10}(\mu)$ is scale-independent.
   The coupling $h_A$ has been measured with great precision by the recent PIBETA experiment
\cite{Bychkov:2008ws}, resulting in \cite{pdg}
\begin{equation}
\left({h_A\over h_V}\right)_{\rm expt}=0.469\pm 0.031
\end{equation}
which then corresponds to the one-loop prediction
\begin{equation}
\alpha_E^{\pi^+}=-\beta_M^{\pi^+}=(2.8\pm 0.2)\times 10^{-4}\,{\rm fm}^3.
\end{equation}

   Two-loop corrections are expected to be small by power-counting arguments,
\begin{equation}
\alpha_E^{\pi^+}|_\text{two-loop}/\alpha_E^{\pi^+}|_\text{one-loop}\sim {4M_\pi^2\over \Lambda_\chi^2}\sim 0.1,
\end{equation}
where $\Lambda_\chi\sim 4\pi F_\pi$ is the chiral-symmetry-breaking scale
\cite{Manohar:1983md,Donoghue:1984gg}.
   This expectation is borne out by evaluation of the Baldin sum rule \cite{bsr} for the charged pion,
which yields \cite{ptr}
\begin{equation}
\alpha_E^{\pi^+}+\beta_M^{\pi^+}
=\frac{1}{2\pi^2}\int_0^\infty d\omega\frac{\sigma_{\rm tot}^{\pi^+}(\omega)}{\omega^2}
=(0.39\pm 0.04)\times 10^{-4}\,{\rm fm}^3
\end{equation}
and by an actual two-loop ChPT calculation \cite{Burgi:1996mm,Burgi:1996qi,Gasser:2006qa}, which, using updated values
for the LECs, yields \cite{Gasser:2006qa}
\begin{equation}
\begin{split}
(\alpha_E^{\pi^+}+\beta_M^{\pi^+})_\text{two-loop}&=0.16\times 10^{-4}\,{\rm fm}^3,\\
(\alpha_E^{\pi^+}-\beta_M^{\pi^+})_\text{two-loop}&=(5.7\pm 1.0)\times 10^{-4}\,{\rm fm}^3.
\end{split}
\end{equation}
  That the relation Eq.~(\ref{eq:alphapiplus})  should exist between the pion Compton amplitude and that for
axial radiative charged-pion beta decay can be understood from simple current
algebra arguments in combination with the hypothesis of a partially conserved axial-vector
current \cite{Terentev:1972ix}---
\begin{equation}
T^{\mu\nu}_{\rm Compton}(p_2,q_2;p_1,q_1)\stackrel{p_2\rightarrow 0}{\longrightarrow}
{i\over \sqrt{2}F_\pi}\left(A^{\mu\nu}(p_1,q_1)+A^{\nu\mu}(p_1,q_2)\right).
\end{equation}
   Thus, on the theoretical side we have a rather precise and solid prediction for
the charged-pion polarizability and we move to experimental tests.

   There have been three different techniques used in the experimental study of the
charged-pion polarizability.
   Since a charged-pion target does not exist, these methods are each indirect and
we consider them in turn:
\begin{itemize}
\item[a)]  The most direct method involves the use of a high-energy pion beam and
the $(\pi^+,\pi^+\gamma)$ reaction.
   Extrapolation to the photon pole using the Primakoff effect then leads to a
measurement of the $\gamma\pi^+\rightarrow\gamma\pi^+$ amplitude and
thereby the charged-pion polarizability.
   An experiment by Antipov et al.~at Dubna yielded the result \cite{Antipov:1982kz,Antipov:1984ez}
\begin{equation}
\alpha_E^{\pi^+}|_\text{expt-a1}=-\beta_M^{\pi^+}|_\text{expt-a1}=(6.8\pm 1.4)\times 10^{-4}\,{\rm fm}^3.
\end{equation}
    A recent experiment by the COMPASS collaboration using the $(\mu^+,\mu^+\gamma)$ reaction as a normalization,
has produced a new preliminary number \cite{Friedrich:2013iya}
\begin{equation}
\alpha_E^{\pi^+}|_\text{expt-a2}=-\beta_M^{\pi^+}|_\text{expt-a2}=(1.9\pm 0.7\pm 0.8)\times 10^{-4}\,{\rm fm}^3.
\end{equation}
\item[b)]
   An alternative way to access the pion Compton amplitude is via the
radiative pion photoproduction reaction $\gamma p\to\gamma n \pi^+$.
   The analysis of a measurement by Aibergenov et al.\ \cite{Aibergenov:1986gi} at the
Lebedev Physical Institute made use of an extrapolation to the pion pole at $t=M_\pi^2$
in the unphysical region, yielding the number
\begin{equation}
\alpha_E^{\pi^+}|_\text{expt-b1}=-\beta_M^{\pi^+}|_\text{expt-b1}=(20\pm 12)\times 10^{-4}\,{\rm fm}^3.
\end{equation}
   A different approach avoiding an extrapolation was chosen in the analysis of the more recent experiment
at the Mainz Microtron (MAMI) \cite{Ahrens:2004mg}.
   The $\pi^+$ polarizabilities have been determined from a comparison of the data
with the predictions of two different theoretical models:
\begin{equation}
\label{alphambeta} (\alpha_E^{\pi^+}-\beta_M^{\pi^+})|_\text{expt-b2}
=(11.6\pm 1.5_{\rm stat}\pm 3.0_{\rm syst}\pm 0.5_{\rm mod}) \times 10^{-4}\, \mbox{fm}^3.
\end{equation}
\item[c)]
   A direct measurement via the $\gamma\gamma\rightarrow\pi^+\pi^-$ reaction at SLAC has yielded the
result \cite{Babusci:1991sk}
\begin{equation}
\alpha_E^{\pi^+}|_\text{expt-c}=-\beta_M^{\pi^+}|_\text{expt-c}=(2.2\pm 1.1)\times 10^{-4}\,{\rm fm}^3.
        \end{equation}
\end{itemize}
   There is thus considerable experimental uncertainty at the present time and additional
experimental work in this regard is urgently needed in order to confirm these presumably
solid chiral perturbative predictions.
 An approved JLab experiment using the Primakoff effect and the $(\gamma,\pi^+\pi^-)$ reaction
should be helpful in this regard \cite{Miskimen}.

\subsection{Neutral-Pion and Kaon Polarizabilities}
   There has also been a substantial amount of work on neutral-pion pair production in
two-photon collisions.
   Data were taken by the Crystal Ball Collaboration for
$\pi^0\pi^0$ invariant masses $W$ from threshold to 2 GeV \cite{Marsiske:1990hx}
and, more recently, by the Belle Collaboration
\cite{Uehara:2008ep} in the kinematic range 0.6 GeV $\leq W\leq $ 4.0 GeV.
   At the one-loop level, ChPT makes a parameter-free prediction for the neutral-pion
polarizabilities \cite{Pervushin:1974kc,Bijnens:1987dc,Donoghue:1988eea}:
\begin{equation}
\alpha_E^{\pi^0}=-\beta_M^{\pi^0}=-\frac{\alpha_{\rm em}}{96\pi^2 F_\pi^2 M_\pi}
=-0.5 \times 10^{-4}\, \mbox{fm}^3.
\end{equation}
   Note in particular that the electric polarizability is negative.
   Two-loop calculations of the $\gamma\gamma\to\pi^0\pi^0$ reaction have been performed in
Refs.\ \cite{Bellucci:1994eb,Bel'kov:1995fj,Gasser:2005ud} and,
using updated values for the LECs, the two-loop results for the sum and
the difference of polarizabilities are given by \cite{Gasser:2005ud}
\begin{equation}
\begin{split}
(\alpha_E^{\pi^0}+\beta_M^{\pi^0})_\text{two-loop}&=(1.1\pm 0.3) \times 10^{-4}\, \mbox{fm}^3,\\
(\alpha_E^{\pi^0}-\beta_M^{\pi^0})_\text{two-loop}&=-(1.9\pm 0.2) \times 10^{-4}\, \mbox{fm}^3.
\end{split}
\end{equation}
   As in the case of the charged pions,
the degeneracy $\alpha_E^{\pi^0}+\beta_M^{\pi^0}=0$ is lifted at the two-loop
level.
   For a discussion of quadrupole polarizabilities see, e.g.,  Refs.\ \cite{Gasser:2005ud,Fil'kov:2005yf}.
   Recent dispersion-theoretical treatments of the $\gamma\gamma\to\pi\pi$ reaction can be
found in \cite{Fil'kov:2005yf,Pasquini:2008ep,GarciaMartin:2010cw,Hoferichter:2011wk}.

   The one-loop predictions for the charged-kaon polarizabilities are determined by
the same linear combination of LECs as those of the charged-pion polarizabilities.
   One simply needs to replace the pion mass and pion-decay constant by the kaon mass
and kaon-decay constant, respectively \cite{Donoghue:1989si}:
\begin{equation}
\alpha_E^{K^+}=-\beta_M^{K^+}=4\frac{\alpha_{\rm em}}{F_K^2 M_K}\left(L_9^r(\mu)+L_{10}^r(\mu)\right)
=0.58 \times 10^{-4}\, \mbox{fm}^3,
\end{equation}
while the neutral-kaon polarizabilities vanish at this order \cite{Guerrero:1997rd}.
   Unfortunately, except for an upper limit $|\alpha_E^{K^-}|<200 \times 10^{-4}\, \mbox{fm}^3$
from kaonic atoms \cite{Backenstoss:1973jx}, no experimental information on kaon polarizabilities
is available, though in principle, the charged-kaon polarizabilities could also be investigated by the
COMPASS collaboration via the Primakoff reaction with a kaon beam \cite{Moinester:2003rb}.

\section{Spin Polarizabilities}
   As emphasized above, in a low-energy expansion the
leading---${\cal O}(\omega^2)$---correction to the Thomson amplitude is given by the
spin-independent (scalar) electric and magnetic polarizability terms given in Eq.~(\ref {eq:gv}).
   At higher---${\cal O}(\omega^3)$---order there exist four additional---{\it spin-dependent}---forms
(spin polarizabilities) as first pointed out by Ragusa \cite{Ragusa:1993rm,Ragusa:1994pm}, which can be written
in the form \cite{Babusci:1998ww}
\begin{equation}
\begin{split}
H^{(3)}_{\rm eff}&=-\frac{1}{2}4\pi
\Big(\gamma_{E1E1}\,\boldsymbol{\sigma}\cdot\mathbf{E}\times\dot{\mathbf{E}}
+\gamma_{M1M1}\,\boldsymbol{\sigma}\cdot\mathbf{H}\times\dot{\mathbf{H}}\\
&\quad+2\gamma_{E1M2}\,\sigma_iE_jH_{ij}-2\gamma_{M1E2}\,\sigma_iH_jE_{ij}\Big),
\end{split}
\label{eq:Heff3}
\end{equation}
where $E_{ij},H_{ij}={1\over 2}(\nabla_iE_j+\nabla_jE_i),{1\over 2}(\nabla_iH_j+\nabla_jH_i)$
are field gradients and the subscripts on the spin polarizabilities indicate the
associated excitation/deexcitation photon multipolarities.
   In this case there is no simple classical analogy in order to understand the significance
of the spin polarizabilities.
   The Compton scattering amplitude can be written in the spin-dependent form
\begin{equation}
\begin{split}
T_{\rm RCS}&=\mathbf{\hat{\boldsymbol\epsilon}}_f^*\cdot\mathbf{\hat{\boldsymbol\epsilon}}_i\, A_1
+\mathbf{\hat{\boldsymbol\epsilon}}_f^*\cdot\mathbf{\hat q}_i\,
\mathbf{\hat{\boldsymbol\epsilon}}_i\cdot\mathbf{\hat q}_f\, A_2
+i\boldsymbol{\sigma}\cdot(\hat{\boldsymbol\epsilon}_f^*\times\hat{\boldsymbol\epsilon}_i)A_3\\
&\quad+i\boldsymbol{\sigma}\cdot(\hat{\mathbf q}_f\times\hat{\mathbf q}_i)
\hat{\boldsymbol\epsilon}_f^*\cdot\hat{\boldsymbol\epsilon}_iA_4
+i{\boldsymbol\sigma}\cdot[(\hat{\boldsymbol\epsilon}_f^*\times\hat{\mathbf q}_i)
\hat{\boldsymbol\epsilon}_i\cdot\hat{\mathbf q}_f
-(\hat{\boldsymbol\epsilon}_i\times\hat{\mathbf q}_f)\hat{\boldsymbol\epsilon}_f^*\cdot\hat{\mathbf q}_i]A_5\\
&\quad+i\boldsymbol{\sigma}\cdot[(\hat{\boldsymbol\epsilon}_f^*\times\hat{\mathbf q}_f)
\hat{\boldsymbol\epsilon}_i\cdot\hat{\mathbf q}_f
-(\hat{\boldsymbol\epsilon}_i\times\hat{\mathbf q}_i)\hat{\boldsymbol\epsilon}_f^*\cdot\hat{\mathbf q}_i]A_6
\end{split}
\end{equation}
with
\begin{equation}
A_i=A_i^\text{Born}+A_i^\text{non-Born}.
\end{equation}
   Here the Born amplitudes are given by (with the superscript $p$ omitted for notational reasons)
\begin{equation}
\begin{split}
\label{equation:AiBorn}
A_1^{\rm Born}&=-\frac{e^2}{m_p},\quad A_2^{\rm Born}=\frac{e^2\omega}{m_p^2},\quad
A_3^{\rm Born}=\frac{e^2\omega}{2m_p^2}\left[1+2\kappa_p-(1+\kappa_p)^2\cos\theta\right],\\
A_4^{\rm Born}&=-A_5^{\rm Born}=-\frac{e^2\omega(1+\kappa_p)^2}{2m_p^2},
\quad A_6^{\rm Born}=-\frac{e^2\omega(1+\kappa_p)}{2m_p^2},
\end{split}
\end{equation}
where $\kappa_p=1.793$ is the proton anomalous magnetic moment in units of the nuclear
magneton.
   Equation (\ref{equation:AiBorn}) agrees with the predictions of the low-energy theorem of
real Compton scattering on a spin-1/2 target \cite{GellMann:1954kc,Low:1954kd},
while the structure-dependent counterparts from Eq.\ (\ref{eq:Heff3}) are
\begin{equation}
\begin{split}
A_1^\text{non-Born}&=4\pi(\alpha_E+\beta_M\cos\theta)\omega^2+{\cal O}(\omega^3),\\
A_2^\text{non-Born}&=-4\pi\beta_M\omega^2+{\cal O}(\omega^3),\\
A_3^\text{non-Born}&=-4\pi[\gamma_{E1E1}+\gamma_{E1M2}+(\gamma_{M1M1}
+\gamma_{M1E2})\cos\theta]\omega^3+{\cal O}(\omega^4),\\
A_4^\text{non-Born}&=-4\pi(\gamma_{M1M1}-\gamma_{M1E2})\omega^3+{\cal O}(\omega^4),\\
A_5^\text{non-Born}&=4\pi\gamma_{M1M1}\omega^3+{\cal O}(\omega^4),\\
A_6^\text{non-Born}&=4\pi\gamma_{E1M2}\omega^3+{\cal O}(\omega^4).
\end{split}
\end{equation}
   However, there exists a significant contribution to each spin polarizability from the
$t$-channel pion-pole diagram which carries no new information about nucleon structure,
\begin{equation}
\gamma_{E1E1}^\text{$\pi$-pole}=-\gamma_{M1M1}^\text{$\pi$-pole}=\gamma_{E1M2}^\text{$\pi$-pole}
=-\gamma_{M1E2}^\text{$\pi$-pole}=\frac{\alpha_{\rm em} g_A}{8\pi^2F_\pi^2M_\pi^2}
=10.7\times 10^{-4}\,\mbox{fm}^4.
\end{equation}
   The structure-dependent pieces of the spin polarizabilities are then given
by subtracting these pole contributions
\begin{equation}
\gamma_i^{\rm structure}=\gamma_i-\gamma_i^\text{$\pi$-pole}.
\end{equation}
   In one-loop chiral perturbation theory we determine the ${\cal O}(q^3)$ predictions
\cite{Bernard:1995dp,Hemmert:1997tj}
\begin{equation}
\begin{split}
\gamma_{E1E1}^{\rm structure}
&=-5\gamma_{M1M1}^{\rm structure}
=5\gamma_{E1M2}^{\rm structure}
=5\gamma_{M1E2}^{\rm structure}\\
&=-\frac{\alpha_{\rm em} g_A^2}{192\pi^2F_\pi^2 M_\pi^2}=-0.569\times 10^{-4}\,\mbox{fm}^4
\end{split}
\end{equation}
which are much smaller than the corresponding pion-pole contributions.
      Full one-loop calculations to ${\cal O}(q^4)$ have been performed in
Refs.~\cite{Gellas:2000mx:6,VijayaKumar:2000pv:6}.
   No new LECs, except for the anomalous magnetic moments
\index{anomalous magnetic moment} of the nucleon,
enter at this order, but the degeneracy between proton and neutron
polarizabilities is lifted.
   Unfortunately, the next-to-leading-order contributions turn out to be
very large, calling the convergence of the expansion into question
\cite{VijayaKumar:2000pv:6}.
   Various theoretical predictions for the spin-dependent polarizabilities
of the proton are summarized in Table \ref{table:spinpol}.
\begin{table}[t]
\caption[]{ Theoretical predictions for the {\it structure-dependent}
contribution to the spin polarizabilities of the proton
to ${\mathcal{O}}(q^3)$ in HBChPT, to
${\mathcal{O}}(q^4)$ in HBChPT from two derivations, to $
{\mathcal{O}}(\varepsilon^3)$ in the small-scale expansion, in fixed-$t$
dispersion relation analyses (BGLMN) and (HDPV), in a calculation with hyperbolic
dispersion relations (HYP. DR) at $\theta_{\mathrm{lab}}=180^{\mathrm{o}}$,
in a dressed $K$-matrix model (KS), and from chiral dynamics with unitarity and
causality (GLP). The values are given in units of $10^{-4}$ fm$^4$.
Here $\gamma_0^p$ and $\gamma_\pi^p$ are the combinations relevant for forward
and backward scattering and are defined in Eqs.\ (\ref{eq:gamma0}) and (\ref{eq:gammapi}).
\label{table:spinpol}} \vspace{0.5 truecm}
\begin{center}
\renewcommand{\arraystretch}{1.3}
\begin{tabular}{lcccccc}
\hline\hline
&$\gamma_{E1E1}^p$&$\gamma_{M1M1}^p$&$ \gamma_{E1M2}^p$&$\gamma_{M1E2}^p$&$\gamma_0^p$&$\gamma_\pi^{p}$\\
\hline
${\mathcal{O}}(q^3)$ \cite{Bernard:1995dp,Hemmert:1997tj}&$-5.7$&$-1.1$&$1.1$&$1.1$&$4.6$&$4.6$\\
${\mathcal{O}}(q^4)$ \cite{Gellas:2000mx:6}&$-1.8$&$0.4$&$0.7$&$1.8$&$-1.1$&$3.3$\\
${\mathcal{O}}(\varepsilon^3)$ \cite{Gellas:2000mx:6}&$-5.4$&$1.4$&$1.0$&$1.0$&$2.0$&$6.8$\\
${\mathcal{O}}(q^4)$ \cite{VijayaKumar:2000pv:6}&$-1.4$&$3.3$&$0.2$&$1.8$&$-3.9$&$6.3$\\
BGLMN \cite{Babusci:1998ww}&$-3.4$&$2.7$&$0.3$&$1.9$&$-1.5$&$7.8$\\
HDPV \cite{Holstein:1999uu}&$-4.3$&$2.9$&$-0.01$&$2.1$&$-0.7$&$9.3$\\
HYP.~DR \cite{Drechsel:2002ar}&$-3.8$&$2.9$&$0.5$&$1.6$&$-1.1$&$7.8$\\
KS \cite{Kondratyuk:2001qu}&$-5.0$&$3.4$&$-1.8$&$1.1$&$2.4$&$11.4$\\
GLP \cite{Gasparyan:2011yw}&$-3.7$&$2.5$&$1.2$&$1.2$&$-1.2$&$6.1$\\
\hline
\hline
\end{tabular}
\end{center}
\end{table}

   At the present time there exists no direct determination of the four spin polarizabilities
from Compton scattering, though there are existing programs for such measurements at MAMI and at the
High Intensity Gamma-Ray Source (HIGS) at Duke University.
   However, there are two different determinations of various combinations, one which
is relevant in the case of forward Compton scattering and the second which applies
to backward scattering.

   We begin with the forward case.
   In order to see how this constraint comes about consider the general amplitude
for forward Compton scattering from the nucleon---Eq.\ (\ref{eq:bh}).
   Here the dispersion relation involving the spin-independent amplitude $f_0(\omega)$ was given above,
while that for the corresponding spin-dependent amplitude $g_0(\omega)$ is \cite{GellMann:1954db}
\begin{equation}
{\rm Re}\left[g_0(\omega)\right]=\frac{\omega}{4\pi^2}P\int_0^\infty d\omega'
\frac{\omega'}{\omega'^2-\omega^2}\left(\sigma_{1/2}(\omega')-\sigma_{3/2}(\omega')\right),
\end{equation}
where $\sigma_{3/2}(\omega'),\sigma_{1/2}(\omega')$ are the photoabsorption cross sections
in the case that the incident photon helicity is parallel, antiparallel to the target spin, respectively,
and, since there is a spin flip involved, the dispersion relation does {\it not} require a subtraction.
   The spin-dependent component of the proton forward Compton amplitude is given by
\begin{equation}
g_0^p(\omega)=-{e^2\kappa_p^2\over 8\pi m_p^2}\omega+\gamma_0^p\omega^3+\ldots,
\end{equation}
where
\begin{equation}
\gamma_0=-\gamma_{E1E1}-\gamma_{M1M1}-\gamma_{E1M2}-\gamma_{M1E2}
\label{eq:gamma0}
\end{equation}
is the combination of spin polarizabilities relevant for forward Compton scattering.
   Note that the pion-pole contribution cancels out in the forward direction, so that only
structure-dependent components remain.
   Equating the two forms for the spin-flip amplitude we find two sum rules.
   One is for the anomalous magnetic moment and is the well-known
Gerasimov-Drell-Hearn (GDH) sum rule \cite{Gerasimov:1965et,Drell:1966jv}
\begin{equation}
\frac{2\pi^2\alpha_{\rm em}\kappa_p^2}{m_p^2}
=\int_0^\infty \frac{d\omega}{\omega}\left(\sigma_{3/2}^p(\omega)
-\sigma_{1/2}^p(\omega)\right)\equiv I_{\rm GDH}^p
\label{eq:hy}
\end{equation}
and the second is a sum rule for the forward spin polarizability
\begin{equation}
\gamma_0^p=-\frac{1}{4\pi^2}\int_0^\infty \frac{d\omega}{\omega^3}
\left(\sigma_{3/2}^p(\omega)-\sigma_{1/2}^p(\omega)\right).
\end{equation}
   The numerical value of the LHS of Eq.\ (\ref{eq:hy}) is 204.8 $\mu$b.
   The GDH collaboration has measured the cross section difference
$(\sigma^p_{3/2}-\sigma^p_{1/2})$ in the photon energy
range from 0.2 to 2.9 GeV with the tagged photon facilities at MAMI (Mainz)
and ELSA (Bonn) \cite{Ahrens:2000bc,Ahrens:2001qt,Dutz:2003mm,Dutz:2004zz}.
   Using theoretical input for the energy regions below $\omega=0.2$ GeV and
above $\omega=2.9$ GeV, the total RHS result was determined to be \cite{Drechsel:2007sq}
(see Table \ref{table_GDH})
\begin{equation}
I_{\rm GDH}^p=(211\pm 15)\,\mu{\rm b},
\end{equation}
so that the proton GDH sum rule is satisfied.
\begin{table}
\centering \caption{Contribution of various energy regions to
the GDH integral $I_{\rm GDH}^p$ and the forward spin polarizability
$\gamma_0^p$ of the proton \label{table_GDH}}
\vspace{0.3cm}
\renewcommand{\arraystretch}{1.3}
\begin{tabular}{@{}lcc@{}}
Energy range & $I^p_{\rm GDH}$ $[\mu$b]  & $\gamma_0^p$ $[10^{-4}$~fm$^4]$ \\
\hline
\hline
$\omega\le200$~MeV~\cite{Drechsel:1998hk,Arndt:2002xv}  &  $-28.5\pm2$  &  $0.95\pm0.05$\\
200 MeV$\le\omega\le800$~MeV~\cite{Ahrens:2001qt}  &  $226\pm5\pm12$  &  $-1.87\pm0.08\pm0.10$\\
800 MeV$\le\omega\le2.9$~GeV~\cite{Dutz:2004zz}  &  $27.5\pm2.0\pm1.2$  & $-0.03$  \\
$\omega\ge2.9$~GeV~\cite{Bianchi:1999qs,Simula:2001iy}  &  $-14\pm2$  &  $+0.01$ \\
\hline
Total & $211\pm15$  &  $-0.94\pm0.15$ \\
\hline
Sum rule~\cite{Gerasimov:1965et,Drell:1966jv} &  204.8  &  -- \\
\hline
\hline
\end{tabular}
\end{table}
   The sum rule for the forward spin polarizability then becomes \cite{Drechsel:2007sq}
(see Table \ref{table_GDH})
\begin{equation}
\gamma_0^\text{$p$-sum-rule}=(-0.94\pm 0.15)\times 10^{-4}\,{\rm fm}^4
\end{equation}
which is in strong disagreement with the one-loop HBChPT prediction---
\begin{equation}
\gamma_0^\text{$p$-one-loop}={\alpha_{\rm em} g_A^2\over 24\pi^2F_\pi^2M_\pi^2}=4.55\times 10^{-4}\,{\rm fm}^4
\end{equation}
so that there must exist a significant contribution from higher-order terms.

   In the backward direction there exists a different combination of spin polarizabilities
which is relevant
\begin{equation}
\gamma_\pi=-\gamma_{E1E1}+\gamma_{M1M1}-\gamma_{E1M2}+\gamma_{M1E2}
\label{eq:gammapi}
\end{equation}
and this backward spin polarizability has been determined experimentally from a
global fit to Compton scattering \cite{Schumacher:2005an}
\begin{equation}
\gamma_\pi^\text{$p$-exp}=-(38.7\pm 1.8)\times 10^{-4}\,{\rm fm}^4.
\label{eq:jy}
\end{equation}
   Note that in this case the pion pole {\it does} contribute
\begin{equation}
\gamma_\pi^\text{$p$-$\pi^0$-pole}
=-{\alpha_{\rm em} g_A\over 2\pi^2F_\pi^2M_\pi^2}=-42.8\times 10^{-4}\,{\rm fm}^{-4}.
\label{eq:fe}
\end{equation}
   The difference between Eqs.\ (\ref{eq:jy}) and (\ref{eq:fe}) is then the
structure component of the backward spin polarizability and is much smaller than the
pion-pole piece
\begin{equation}
\gamma_\pi^\text{$p$-structure}=(4.0\pm 1.8)\times 10^{-4}\,{\rm fm}^4.
\end{equation}
   In this case the one-loop---${\cal O}(q^3)$---chiral prediction is found to be
\begin{equation}
\gamma_\pi^\text{$p$-one-loop}={\alpha_{\rm em} g_A^2\over 160\pi^2F_\pi^2M_\pi^2}=0.683\times 10^{-4}\,{\rm fm}^4.
\end{equation}

   At the present time, the only reliable data which is available is that for the forward and backward
spin polarizabilities.
   However, in the near future this situation will change substantially due to Compton scattering
experiments involving polarized photons and polarized targets.
   This effort should take place at MAMI, where such work has already begun and at
HIGS, where improved mirrors should enable a viable future program.
   At MAMI, a series of measurements is envisioned in the photon energy range of about 80-300 MeV
and is of three types:
\begin{itemize}
\item[a)] Linearly polarized photons either parallel or perpendicular to the scattering plane
with an unpolarized target (usually called the beam asymmetry):
\begin{equation}
\Sigma_3={\sigma_\parallel-\sigma_\perp\over \sigma_\parallel+\sigma_\perp}.
\end{equation}
\item[b)] Circularly polarized photons and target spin aligned longitudinally with the beam direction:
\begin{equation}
\Sigma_{2z}={\sigma_{+z}^R-\sigma_{+z}^L\over \sigma_{+z}^R+\sigma_{+z}^L}
={\sigma_{+z}^R-\sigma_{-z}^R\over \sigma_{+z}^R+\sigma_{-z}^R}.
\end{equation}
\item[c)] Circularly polarized photons and target spin aligned transverse to the beam direction:
\begin{equation}
\Sigma_{2x}={\sigma_{+x}^R-\sigma_{+x}^L\over \sigma_{+x}^R
+\sigma_{+x}^L}={\sigma_{+x}^R-\sigma_{-x}^R\over \sigma_{+x}^R+\sigma_{-x}^R}.
\end{equation}
\end{itemize}
   Here the measurements of $\Sigma_{2x}$  and $\Sigma_3$ have already been accomplished, while the
remaining measurement of $\Sigma_{2z}$ is scheduled to take place in  2014 \cite{Arends}.
   Analysis of the $\Sigma_{2x}$ data has already led to a preliminary value for
$\gamma_{E1E1}^p=(-4.3\pm 1.5)\times 10^{-4}\,{\rm fm}^4$ \cite{Middleton:2013mya}, the data on $\Sigma_3$
are presently analyzed, and the remaining data should enable determination of the other
spin polarizabilities.

\section{Higher-Order Polarizabilities}
   It is obvious that the concept of polarizabilities can be generalized to
even higher orders \cite{Holstein:1999uu} (see also Refs.~\cite{Griesshammer:2001uw,Hildebrandt:2003fm}
for a definition of so-called dynamical polarizabilities).
   For example, in the spin-independent sector the ${\cal O}(\omega^4)$ generalization of the effective
Hamiltonian can be written as \cite{Babusci:1998ww}
\begin{equation}
H_{\rm eff}^{(4)}=-\frac{1}{2}4\pi\alpha_{E\nu}{\dot{\mathbf{E}}}^2
-{1\over 2}4\pi\beta_{M\nu}{\dot{\mathbf{H}}}^2
-{1\over 12}4\pi\alpha_{E2}E_{ij}^2
-{1\over 12}4\pi\beta_{M2}H_{ij}^2.
\end{equation}
   The meaning of these new quadratic terms is clear.
   The quantities $\alpha_{E\nu}$ and $\beta_{M\nu}$ represent dispersive corrections to the
leading electric and magnetic polarizabilities and measure the frequency dependence of the
electric and magnetic polarizabilities via
\begin{equation}
\begin{split}
\mathbf{p}(\omega)&=4\pi(\alpha_E+\alpha_{E\nu}\omega^2+\ldots)\mathbf{E}(\omega),\\
\mathbf{m}(\omega)&=4\pi(\beta_M+\beta_{M\nu}\omega^2+\ldots)\mathbf{H}(\omega).
\end{split}
\end{equation}
   Likewise, it is clear that the quadrupole polarizabilities $\alpha_{E2}$ and $\beta_{M2}$
measure the size of the induced quadrupole moments in the presence of an applied field gradient via
\begin{equation}
\begin{split}
Q_{ij}&=-{\partial H_{\rm eff}^{(4)}\over \partial E_{ij}}={1\over 6}4\pi\alpha_{E2}E_{ij},\\
M_{ij}&=-{\partial H_{\rm eff}^{(4)}\over \partial H_{ij}}={1\over 6}4\pi\alpha_{M2}H_{ij},
\end{split}
\end{equation}
where
\begin{equation}
Q_{ij}=\big\langle\psi\big|\sum_{k=1}^3 q_k\left[3(r_k-R)_i(r_k-R)_j
-\delta_{ij}(\mathbf{r}_k-\mathbf{R})^2\right]\big|\psi\big\rangle
\end{equation}
is the induced electric quadrupole moment and $M_{ij}$ is its magnetic analog.

   These four new polarizabilities can be calculated in various pictures of the nucleon.
For example, in the simple harmonic oscillator picture the sum rules
\begin{equation}
\alpha_{E\nu}^p=2\alpha_{\rm em}\sum_{n\neq 0}\frac{|\langle n|D_z|0\rangle_p|^2}{(E_n-E_0)^3},
\quad \alpha_{E2}^p=\frac{\alpha_{\rm em}}{2}\sum_{n\neq 0}\frac{|\langle n|Q_{33}|0\rangle_p|^2}
{E_n-E_0}
\end{equation}
yield the predictions
\begin{equation}
\alpha_{E\nu}^p=\frac{2\alpha_{\rm em} m_p^3}{81}\langle (r_E^p)^2\rangle^4,\quad
\alpha_{E2}^p=\frac{\alpha_{\rm em} m_p}{9}\langle (r_E^p)^2\rangle^3
\end{equation}
and similarly the magnetic polarizabilities can be estimated.
   Introducing
$$\xi_1=\frac{\alpha_{\rm em}g_A^2}{1920\,\pi F_\pi^2 M_\pi^3}=2.53\times 10^{-5}\,\mbox{fm}^5,$$
the one-loop $[{\cal O}(q^3)]$ HBChPT predictions of these quantities for the
proton/neutron read \cite{Hemmert:1997tj}
\begin{equation}
\alpha_{E\nu}=9\,\xi_1,\quad \beta_{M\nu}=14\,\xi_1,\quad
\alpha_{E2}=84\,\xi_1,\quad \beta_{M2}=-36\,\xi_1
\end{equation}
and the $\Delta$-resonance corrections have also been given
in the small-scale expansion \cite{Holstein:1999uu}.

   Likewise, in the case that spin dependence is considered, there exist eight new
spin polarizabilities which arise at ${\cal O}(\omega^5)$---
\begin{equation}
\begin{split}
H_{\rm eff}^{(5)}&=-{1\over 2}4\pi\Big[\gamma_{E1\nu}\boldsymbol{\sigma}
\cdot\dot{\mathbf{E}}\times\ddot{\mathbf{E}}
+\gamma_{M1\nu}\boldsymbol{\sigma}\cdot\dot{\mathbf{H}}\times\ddot{\mathbf{H}}\\
&\quad-2\gamma_{E2\nu}\sigma_i\dot{E}_{ij}\dot{H}_j+2\gamma_{M2\nu}\sigma_i\dot{H}_{ij}\dot{E}_j\\
&\quad+4\gamma_{ET}\epsilon_{ijk}\sigma_iE_{jl}\dot{E}_{kl}+4\gamma_{MT}\epsilon_{ijk}\sigma_i\dot{H}_{jl}H_{kl}\\
&\quad-6\gamma_{E3}\sigma_iE_{ijk}H_{jk}+6\gamma_{M3}\sigma_iH_{ijk}E_{jk}\Big],
\end{split}
\end{equation}
where
\begin{equation}
E_{ijk}={1\over 3}(\nabla_i\nabla_j E_k+\nabla_i\nabla_kE_j+\nabla_j\nabla_kE_i)
-{1\over 15}(\delta_{ij}\mathbf{\nabla}^2E_k+\delta_{ik}\mathbf{\nabla}^2E_j+\delta_{jk}\mathbf{\nabla}^2E_i)
\end{equation}
with a corresponding expression for $H_{ijk}$.
   Again we can identify the one-loop heavy-baryon chiral predictions \cite{Holstein:1999uu}
\begin{equation}
\begin{split}
\gamma_{E3}&=20\,\xi_2,\quad
\gamma_{M3}=20\,\xi_2,\quad
\gamma_{ET}=-65\,\xi_2,\quad
\gamma_{MT}=-5\xi_2,\\
\gamma_{E1\nu}&=-945\,\xi_2,\quad
\gamma_{M1\nu}=-45\,\xi_2,\quad
\gamma_{E2\nu}=78\,\xi_2,\quad
\gamma_{M2\nu}=-42\,\xi_2,
\end{split}
\end{equation}
where
$$\xi_2={\alpha_{\rm em} g_A^2\over 43200\,\pi^2F_\pi^2M_\pi^4}=5.05\times 10^{-7}\,\mbox{fm}^6.
$$
   However, again there is an anomaly contribution which should be subtracted
(added) in the proton (neutron) case from the measured value to yield the structure dependence
\begin{equation}
\begin{split}
\gamma_{E3}^\text{$\pi$-pole}&=-10\,\xi_3,\quad
\gamma_{M3}^\text{$\pi$-pole}=10\,\xi_3,\quad
\gamma_{ET}^\text{$\pi$-pole}=5\,\xi_3,\quad
\gamma_{MT}^\text{$\pi$-pole}=-5\,\xi_3,\\
\gamma_{E1\nu}^\text{$\pi$-pole}&=-45\,\xi_3,\quad
\gamma_{M1\nu}^\text{$\pi$-pole}=45\,\xi_3,\quad
\gamma_{E2\nu}^\text{$\pi$-pole}=46\,\xi_3,\quad
\gamma_{M2\nu}^\text{$\pi$-pole}=-46\,\xi_3,
\end{split}
\end{equation}
where
\begin{displaymath}
\xi_3={\alpha_{\rm em} g_A\over 120\,\pi^2F_\pi^2M_\pi^4}=1.43\times 10^{-4}\,\mbox{fm}^6.
\end{displaymath}

   There exist two reliable determinations of combinations of these quantities, which can
be compared directly with experiment.
   These arise from the use of forward dispersion relations.
   In the case of the spin-independent polarizabilities we find \cite{Babusci:1998ww}
\begin{equation}
\begin{split}
\alpha_{E\nu}^p+\beta_{M\nu}^p+
\frac{1}{12}\left(\alpha_{E2}^p+\beta_{M2}^p\right)
&=\frac{1}{4\pi^2}\int_0^\infty \frac{d\omega}{\omega^4}
\left(\sigma_{3/2}^p(\omega)+\sigma_{1/2}^p(\omega)\right)\\
&=5.73\times 10^{-4}\,{\rm fm}^5
\end{split}
\end{equation}
to be compared with the ${\cal O}(q^3)$ chiral prediction
\begin{equation}
\left[\alpha_{E\nu}^p+\beta_{M\nu}^p+\frac{1}{12}\left(\alpha_{E2}^p+\beta_{M2}^p\right)\right]_{{\rm HB},q^3}
=27\,\zeta_1=6.83\times 10^{-4}\,{\rm fm}^5,
\end{equation}
while in the case of the spin-polarizabilities, we have \cite{Pasquini:2010zr}
\begin{align}
\lefteqn{\hspace{-5em}-\left(\gamma_{E1\nu}^p+\gamma_{M1\nu}^p+\gamma_{E2\nu}^p+\gamma_{M2\nu}^p
+\frac{8}{5}\gamma_{E3}^p+\frac{8}{5}\gamma_{M3}^p
+\gamma_{ET}^p+\gamma_{MT}^p\right)}\nonumber\\
&=-\frac{1}{4\pi^2}\int_0^\infty\frac{d\omega}{\omega^5}
\left(\sigma_{3/2}^p(\omega)-\sigma_{1/ 2}^p(\omega)\right)
\nonumber\\
&=(0.42\pm0.09\pm0.09)\times 10^{-4}\,{\rm fm}^6
\end{align}
to be compared with the ${\cal O}(q^3)$ chiral prediction
\begin{align}
\lefteqn{\hspace{-7em}-\left(\gamma_{E1\nu}^p+\gamma_{M1\nu}^p+\gamma_{E2\nu}^p+\gamma_{M2\nu}^p
+\frac{8}{5}\gamma_{E3}^p+\frac{8}{5}\gamma_{M3}^p
+\gamma_{ET}^p+\gamma_{MT}^p\right)_{{\rm HB},q^3}}\nonumber\\
&=960\,\zeta_2=4.85 \times 10^{-4}\,{\rm fm}^6.
\end{align}
   However, separating the various polarizabilities is more difficult.
   While in principle the individual higher-order polarizabilities can be determined
from analysis of experimental data, in practice this is not feasible.
   The problem is that the effective action picture is an expansion in powers of
$\omega/\Lambda_\chi$, where  $\Lambda_\chi\sim 4\pi F_\pi$ is the chiral scale.
   As discussed above, at the lowest energies, the cross section is determined
simply by the Born terms and is given by the Powell cross section.
   In the region 50 MeV$\leq \omega\leq$ 100 MeV the dipole polarizabilities come into play
and it is this region which has been used in the experimental determination of these quantities.
   Even at these relatively low energies, however, there are important contributions from the higher-order
terms and these must be estimated (usually dispersively) in order to perform the experimental
extraction of $\alpha_E^p,\beta_M^p$.
   This problem is exacerbated at higher energies so that it is not realistic to think that
one can extract quadrupole polarizabilities or higher in this fashion.

   It {\it is}, however, possible to make an experimental determination by {\it indirect} means.
   That is, one can use subtracted fixed-$t$ dispersion relations to provide a complete low-energy
analysis of the Compton amplitude and then to extract the relevant polarizabilities
from this calculated amplitude.
   In this approach, one writes each invariant amplitude $A_i(\nu,t)$ in the form \cite{pas}
\begin{equation}
{\rm Re}[A_i(\nu,t)]
=A_i^{\rm Born}(\nu,t)+[A_i(0,t)-A_i^{\rm Born}(0,t)]
+\frac{2\nu^2}{\pi}P
\int_{\nu_0}^\infty d\nu'\frac{{\rm Im}_s[A_i(\nu',t)]}{\nu'({\nu'}^2-\nu^2)},
\label{eq:gh}
\end{equation}
where $\nu=(s-u)/4m_N$ is the average of the incoming and outgoing photon
energy---$\nu={1\over 2}(E_\gamma+E_\gamma')$---and ${\rm Im}_s[A_i]$ denotes
the discontinuities across the $s$-channel cuts of the Compton process.
   Because of the three powers of $\nu'$ in the denominator, these subtracted
dispersion relations should converge and moreover, they should be dominated by the
contribution of the $\pi N$ intermediate state, which can be well-determined by experiment.
   Contributions from states with more than a single pion are expected to be small and can be
estimated in simple models.
   The one unknown in Eq.\ (\ref{eq:gh}) is the value $A_i(0,t)$ which is determined via a subtracted
dispersion relation in $t$,
\begin{equation}
\begin{split}
A_i(0,t)&=A_i^{\rm Born}(0,t)+[A_i(0,0)-A_i^{\rm Born}(0,0)]
+[A_i^\text{$t$-pole}(0,t)-A_i^\text{$t$-pole}(0,0)]\\
&\quad+\frac{t}{\pi}\int_{4M_\pi^2}^\infty dt'\frac{{\rm Im}_t[A_i(0,t')]}{t'(t'-t)}
-\frac{t}{\pi}\int_{-\infty}^{-2M_\pi^2-4M_\pi m_N}dt'\frac{{\rm Im}_t[A_i(0,t')]}{t'(t'-t)},
\end{split}
\end{equation}
where $A_i^\text{$t$-pole}(0,t)$ represents the contribution from $t$-channel poles,
including the $\pi^0$ pole that we have discussed above.
   The imaginary component of the amplitudes $A_i(0,t)$ in the integral from $4M_\pi^2$ to $\infty$
is estimated by assuming saturation by light intermediate states such as
$\pi\pi$ and $K\bar{K}$, while that in the integral from $-\infty$ to $-2M_\pi^2-4M_\pi m_N$
is estimated by assuming saturation by the $\Delta$ resonance plus small
nonresonant $\pi N$ contributions.
   In this fashion, the six invariant amplitudes have been determined in terms of the
six subtraction constants $A_i(0,0)$.
   Four of these constants are evaluated via unsubtracted $t=0$ dispersion relations
\begin{equation}
A_i(0,0)=\frac{2}{\pi}\int_{\nu_0}^\infty d\nu\frac{{\rm Im}[A_i(\nu,0)]}{\nu},\quad i=3,4,5,6,
\end{equation}
and the remaining two constants are determined from experiment,
in terms of the constants $(\alpha_E^p-\beta_M^p)$ and $\gamma_\pi^p$.
   With the Compton amplitude now determined, the various polarizabilities can now be found
and compared to chiral predictions, as shown below.
   In the case of the quadrupole spin-independent polarizabilities, we find \cite{Holstein:1999uu}
\begin{equation}
\alpha_{E2}^p=29.31\times 10^{-4}\,{\rm fm}^5,\quad\beta_{M2}^p=-24.33\times 10^{-4}\,{\rm fm}^5
\end{equation}
which can be compared to the ${\cal O}(q^3)$ heavy-baryon chiral predictions \cite{Holstein:1999uu}
\begin{equation}
\alpha_{E2,{\rm HB},q^3}^p=22.1\times 10^{-4}\,{\rm fm}^5,\quad
\beta_{M2,{\rm HB},q^3}^p=-9.5\times 10^{-4}\,{\rm fm}^5.
\end{equation}
   In the case of the higher-order spin polarizabilities, the dispersion-theoretical treatment
yields \cite{Holstein:1999uu}
\begin{equation}
\begin{split}
\gamma_{ET}^p&=-0.15\times 10^{-4}\,{\rm fm}^6,\quad\gamma_{MT}^p=-0.09\times 10^{-4}\,{\rm fm}^6,\\
\gamma_{M3}^p&=0.09\times 10^{-4}\,{\rm fm}^6,\quad\gamma_{E3}^p=0.06\times 10^{-4}\,{\rm fm}^6,
\end{split}
\end{equation}
which can be compared to the ${\cal O}(q^3)$ heavy-baryon chiral predictions \cite{Holstein:1999uu}
\begin{equation}
\begin{split}
\gamma_{ET,{\rm HB},q^3}^p&=-0.37\times 10^{-4}\,{\rm fm}^6,\quad
\gamma_{MT,{\rm HB},q^3}^p=-0.03\times 10^{-4}\,{\rm fm}^6\\
\gamma_{M3,{\rm HB},q^3}^p&=0.11\times 10^{-4}\,{\rm fm}^6,\quad
\gamma_{E3,{\rm HB},q^3}^p=0.11\times 10^{-4}\,{\rm fm}^6.
\end{split}
\end{equation}
We see then that the chiral predictions are generally in the right ballpark with the pattern
of higher-order polarizabilities found via the dispersion-theoretical analysis,
though certainly higher-order contributions are required to obtain real agreement.

\section{Generalized Polarizabilities}
   As in all studies with electromagnetic probes, the possibilities to investigate
the structure of the target are much greater if virtual photons are used,
since the energy and three-momentum of the virtual photon can be varied independently.
   Moreover, the longitudinal component of the current operators entering the amplitude can
be studied.
   The amplitude for virtual Compton scattering (VCS) off the proton, $T_{\rm VCS}^p$,
is accessible in the reaction $e^-(k_i)+p(p_i)\rightarrow\gamma(q')+e^-(k_f)+p(p_f)$.
   In the one-photon-exchange approximation, the scattering amplitude consists of the
Bethe-Heitler (BH) piece, where the real photon is emitted by the initial or final
electrons, and the VCS contribution, $T=T_{\rm BH}+T_{\rm VCS}$ \cite{Berg:1961}.
   The use of a virtual photon in the initial state (four-momentum $q=k_i-k_f$)
means that there are now twelve invariant functions required to describe $T_{\rm VCS}$
\cite{Guichon:1995pu,Scherer:1996ux,Hemmert:1996gr}.
   We shall work in the center-of-mass frame of the final-state photon-nucleon system, where we have
\begin{equation}
\begin{split}
\mathbf{p}_f&=-\mathbf{q}',\\
\mathbf{p}_i&=-\mathbf{q}=-\bar{q}\,\hat{\mathbf e}_z,\\
\omega'+\sqrt{m^2_p+{\omega'}^2}&=\omega+\sqrt{m_p^2+\bar{q}^2},
\end{split}
\end{equation}
with the $z$-axis defined by the three-momentum vector $\mathbf q$ of the incident virtual photon.
   Working in Lorenz gauge
\begin{equation}
\epsilon\cdot q=0,\quad \epsilon_0={\bar{q}\over \omega}\epsilon_z
\end{equation}
with $\boldsymbol{\epsilon}={\boldsymbol\epsilon}_T+\epsilon_z \hat{\mathbf{e}}_z$,
we can represent the VCS transition amplitude as
\begin{align}
T_{\rm VCS}&=\hat{\boldsymbol\epsilon}'^*\cdot\boldsymbol{\epsilon}_T A_1
+\hat{\boldsymbol\epsilon}'^*\cdot\hat{\mathbf q}\,\boldsymbol{\epsilon}_T\cdot\hat{\mathbf q}' A_2
+i\boldsymbol{\sigma}\cdot(\hat{\boldsymbol\epsilon}'^*\times\boldsymbol{\epsilon}_T) A_3
+i\boldsymbol{\sigma}\cdot(\hat{\mathbf q}'\times\hat{\mathbf q})\,
\hat{\boldsymbol\epsilon}'^*\cdot\boldsymbol{\epsilon}_T A_4\nonumber\\
&\quad+i\boldsymbol{\sigma}\cdot(\hat{\boldsymbol\epsilon}'^*\times\hat{\mathbf q})
\boldsymbol{\epsilon}_T\cdot\hat{\mathbf q}'A_5
+i\boldsymbol{\sigma}\cdot(\hat{\boldsymbol\epsilon}'^*\times\hat{\mathbf q}')
\boldsymbol{\epsilon}_T\cdot\hat{\mathbf q}\,A_6\nonumber\\
&\quad-i\boldsymbol{\sigma}\cdot(\boldsymbol{\epsilon}_T\times\hat{\mathbf q}')
\hat{\boldsymbol\epsilon}'^*\cdot\hat{\mathbf q}\,A_7
-i\boldsymbol{\sigma}\cdot(\boldsymbol{\epsilon}_T\times\hat{\mathbf q})
\hat{\boldsymbol\epsilon}'^*\cdot\hat{\mathbf q}\,A_8\nonumber\\
&\quad+{q^2\over \omega^2}\epsilon_z
\left[\hat{\boldsymbol\epsilon}'^*\cdot\hat{\mathbf q}\,A_9
+i\vec{\boldsymbol\sigma}\cdot(\hat{\mathbf q}'\times\hat{\mathbf q})
\hat{\boldsymbol\epsilon}'^*\cdot\hat{\mathbf q}\,A_{10}\right.\nonumber\\
&\left.\quad+i\boldsymbol\sigma\cdot(\hat{\boldsymbol\epsilon}'^*\times\hat{\mathbf q})A_{11}
+i\boldsymbol{\sigma}\cdot(\hat{\boldsymbol\epsilon}'^*\times\hat{\mathbf q}')A_{12}\right].
\end{align}
   Here each amplitude $A_i$ ($i=1,2,\ldots, 12$) is a function of the three kinematic
quantities $\omega',\bar{q},\theta$.

   Model-independent predictions, based on Lorentz invariance, gauge invariance, crossing symmetry,
and the discrete symmetries, have been derived in Ref.\ \cite{Scherer:1996ux}.
   Up to and including terms of second order in $\bar{q}$ and $\omega'$, the amplitude
is completely specified in terms of quantities which can be obtained from other processes,
namely $m_p$, $\kappa_p$, the electric and magnetic Sachs form factors
$G_E^p$ and $G_M^p$, the electric mean square radius $\langle (r_E^p)^2\rangle$,
and the RCS polarizabilities $\alpha_E^p$ and $\beta_M^p$.

   As in the case of real Compton scattering, each invariant amplitude can be written as the sum of
``Born terms'' and structure-dependent components
\begin{equation}
A_i(\omega',\bar{q},\theta)=A_i^{\rm Born}(\omega',\bar{q},\theta)+A_i^{\rm structure}(\omega',\bar{q},\theta).
\end{equation}
Here the Born term consists of the sum of the nucleon and anomaly ($\pi^0$ pole) pieces
\cite{Knoechlein:1997,Hemmert:1999pz}---
\begin{equation}
\begin{split}
A_1^{\rm Born}&=-{e^2\over m_p}+{\cal O}(1/m_p^3),\\
A_2^{\rm Born}&=\bar{q}{e^2\over m_p^2}+{\cal O}(1/m_p^3),\\
A_3^{\rm Born}&=\left[(1+2\kappa_p)\omega'-(1+\kappa_p)^2\bar{q}\cos\theta\right]
\frac{e^2}{2m_p^2}\\
&\quad -{\omega'({\omega'}^2+\bar{q}^2-2\omega'\bar{q}\cos\theta)\over f(\omega',\bar{q},\theta)}
{e^2 g_A \over 8\pi^2F_\pi^2}+{\cal O}(1/m_p^3),\\
A_4^{\rm Born}&=-(1+\kappa_p)^2\bar{q}{e^2\over 2m_p^2}+{\cal O}(1/m_p^3),\\
A_5^{\rm Born}&=(1+\kappa_p)^2\bar{q}{e^2\over 2m_p^2}
-{{\omega'}^2\bar{q}\over f(\omega',\bar{q},\theta)}{e^2g_A\over 8\pi^2F_\pi^2}+{\cal O}(1/m_p^3),\\
A_6^{\rm Born}&=-(1+\kappa_p)\omega'{e^2\over 2m_p^2}
+{{\omega'}^3\over f(\omega',\bar{q},\theta)}{e^2 g_A\over 8\pi^2F_\pi^2}
+{\cal O}(1/m_p^3),\\
A_7^{\rm Born}&=(1+\kappa_p)^2 \omega' {e^2\over 2m_p^2}
-{{\omega'}^2\bar{q}\over f(\omega',\bar{q},\theta)} {e^2 g_A\over 8\pi^2F_\pi^2}+{\cal O}(1/m_p^3),\\
A_8^{\rm Born}&=-(1+\kappa_p){\bar{q}^2\over \omega'}{e^2 \over 2m_p^2}
+{{\omega'}\bar{q}^2\over f(\omega',\bar{q},\theta)}{e^2g_A\over 8\pi^2F_\pi^2}+{\cal O}(1/m_p^3),\\
A_9^{\rm Born}&=-{e^2\over m_p}+{2\omega'\bar{q}\cos\theta+\bar{q}^2\over\omega'}\frac{e^2}{2m_p^2}
+{\cal O}(1/m_p^3),\\
A_{10}^{\rm Born}&=-{{\omega'}^2\bar{q}\over f(\omega',\bar{q},\theta)}
{e^2 g_A\over 8\pi^2F_\pi^2}+{\cal O}(1/m_p^3),\\
A_{11}^{\rm Born}&=(1+2\kappa_p)\omega'\frac{e^2}{2m_p^2}
-{{\omega'}^2(\omega'-\bar{q}\cos\theta)\over f(\omega',\bar{q},\theta)}{e^2g_A\over 8\pi^2F_\pi^2}
+{\cal O}(1/m_p^3),\\
A_{12}^{\rm Born}&=-(1+\kappa_p)\cos\theta\,\omega'{e^2\over 2m_p^2}
-{{\omega'}^2(\bar{q}-\omega'\cos\theta)\over f(\omega',\bar{q},\theta)}
{e^2g_A\over 8\pi^2F_\pi^2}+{\cal O}(1/m_p^3),
\end{split}
\end{equation}
where
\begin{equation}
f(\omega',\bar{q},\theta)=M_\pi^2+{\omega'}^2+\bar{q}^2-2\omega'\bar{q}\cos\theta
\end{equation}
originates from the pion pole term.

   After dividing the amplitude $T_{\rm VCS}$ into a gauge-invariant generalized Born piece $T^{\rm Born}_{\rm VCS}$
and a structure-dependent residual component $T^{\rm R}_{\rm VCS}$, the so-called {\it generalized} polarizabilities
(GPs) of Ref.\ \cite{Guichon:1995pu} result from an analysis of the residual term in terms of electromagnetic
multipoles.
   A restriction to the lowest order, {\it i.e.}, terms linear in $\omega'$, leads to only electric and magnetic
dipole radiation in the final state.
   Parity and angular-momentum selection rules, charge-conjugation symmetry, and particle crossing generate six
independent GPs \cite{Guichon:1995pu,Drechsel:1996ag,Drechsel:1997xv}.
   Predictions for the GPs of the nucleon have been obtained, for example, in the constituent
quark model \cite{Guichon:1995pu,Pasquini:2000ue},
in HBChPT at ${\cal O}(q^3)$
\cite{Hemmert:1996gr,Hemmert:1997at} and ${\cal O}(q^4)$ \cite{Kao:2002cn,Kao:2004us},
as well as in the small-scale expansion at ${\cal O}(q^3)$
\cite{Hemmert:1999pz}.
   The predictions of HBChPT at ${\cal O}(q^3)$ contain no unknown LECs, {\it i.e.}, they are given in terms of
the pion mass, the axial-vector coupling constant $g_A$, and the pion-decay constant $F_\pi$.
   In the case of the spin-independent polarizabilities we have for the proton and the neutron
\begin{equation}
\begin{split}
\alpha_E(\bar{q}^2)&={5e^2g_A^2\over 384\pi^2F_\pi^2M_\pi}
\left(1-{7\over 50}{\bar{q}^2\over M_\pi^2}+{81\over 2800}{\bar{q}^4\over M_\pi^4}+\ldots\right),\\
\beta_M(\bar{q}^2)&={e^2g_A^2\over 768\pi^2F_\pi^2M_\pi}\left(1+{1\over 5}{\bar{q}^2\over M_\pi^2}
-{39\over 560}{\bar{q}^4\over M_\pi^4}+\ldots\right).
\end{split}
\end{equation}
   Note that at $\bar{q}=0$ the values of the generalized polarizabilities
coincide with their real photon counterparts.
   In the case of the electric polarizability, the $\bar{q}^2$ dependence involves a general fall-off
with the scale $\bar{q}^2/M_\pi^2$ as expected from the contribution of the pion cloud.
   After Fourier transforming we find that the local polarizability involves a maximum at the center and a
falloff with size $\sim\delta\sim 1/M_\pi$.
   In the case of the magnetic polarizability, there is an {\it increase} in momentum space before a
general falloff as expected from the pion-cloud contribution.
   Thus, there is a prediction of both paramagnetic and diamagnetic behavior.

\begin{figure}
\begin{center}
\resizebox{0.6\textwidth}{!}{
\includegraphics{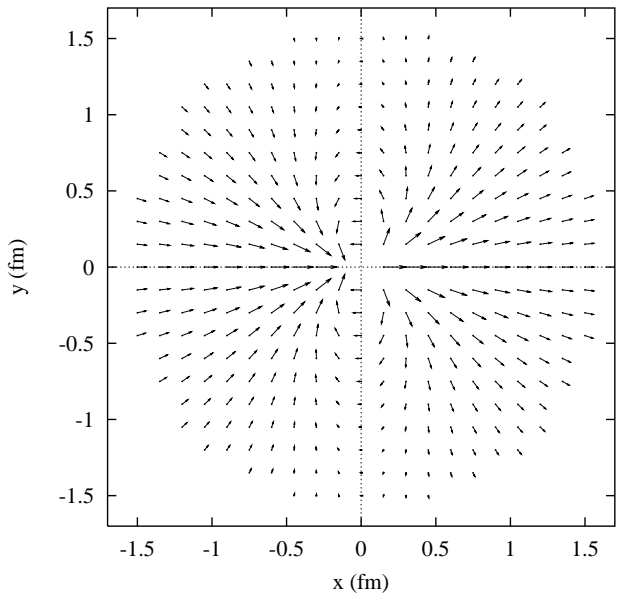}
}
\end{center}
\caption{Scaled electric polarization $r^3 \alpha_{i1}$ [10$^{-3}$ fm$^3$] \cite{Lvov:2004}.
The applied electric field points in the $x$ direction.} \label{fig:polarization}
\end{figure}

   A covariant definition of the spin-averaged dipole polarizabilities was proposed in
Ref.~\cite{L'vov:2001fz}.
   It was shown that {\it three} generalized dipole polarizabilities are needed to reconstruct
the location of the polarization by Fourier transforming.
   For example, if the nucleon is exposed to a static and uniform external electric field  $\mathbf{E}$,
an electric polarization $\boldsymbol{\cal P}$ is generated which is related to the {\it density} of the
induced electric dipole moments,
\begin{equation}
\label{H1:b:d-induced}
{\cal P}_i(\mathbf{r}) = 4\pi\alpha_{ij}(\mathbf{r})\,E_j.
\end{equation}
The tensor $\alpha_{ij}(\mathbf r)$,  {\it i.e.}~the density of the full electric polarizability
of the system, can be expressed as \cite{L'vov:2001fz}
\begin{displaymath}
\alpha_{ij}(\mathbf{r)} =
\alpha_L(r) \frac{r_i r_j}{r^2}
+ \alpha_T(r)\frac{r^2\delta_{ij} - r_i r_j}{r^2}
+ \frac{3r_i r_j - r^2\delta_{ij}}{r^5}
 \int_r^\infty dr' r'^2[\alpha_L(r')-\alpha_T(r')]\,,
\end{displaymath}
where $\alpha_L(r)$ and $\alpha_T(r)$ are Fourier transforms of the generalized longitudinal and
transverse electric polarizabilities $\alpha_L(q^2)$ and $\alpha_T(q^2)$, respectively.
   In particular, it is important to realize that both longitudinal and transverse polarizabilities
are needed to fully recover the electric polarization $\boldsymbol{\cal P}$.
   Figure \ref{fig:polarization} shows the induced polarization inside a proton as calculated
in the framework of HBChPT at ${\cal O}(q^3)$ \cite{Lvov:2004}.
   Since $\alpha_L(q^2)$ and $\alpha_T(q^2)$ differ,
the polarization does not necessarily point into the direction of the applied electric field.
   Similar considerations apply to an external magnetic field.
   Since the magnetic induction is always transverse ({\it i.e.}, $\boldsymbol\nabla\cdot\mathbf B=0$),
it is sufficient to consider $\beta_{ij}(\mathbf r)=\beta(r)\delta_{ij}$ \cite{L'vov:2001fz}.
   The induced magnetization $\boldsymbol{\cal M}$ is given in terms of the density of the magnetic
polarizability as $\boldsymbol{\cal M}(\mathbf r) = 4\pi\beta(r)\mathbf H$.

\begin{figure}
\begin{minipage}[b]{0.5\textwidth}
\begin{center}
\epsfig{file=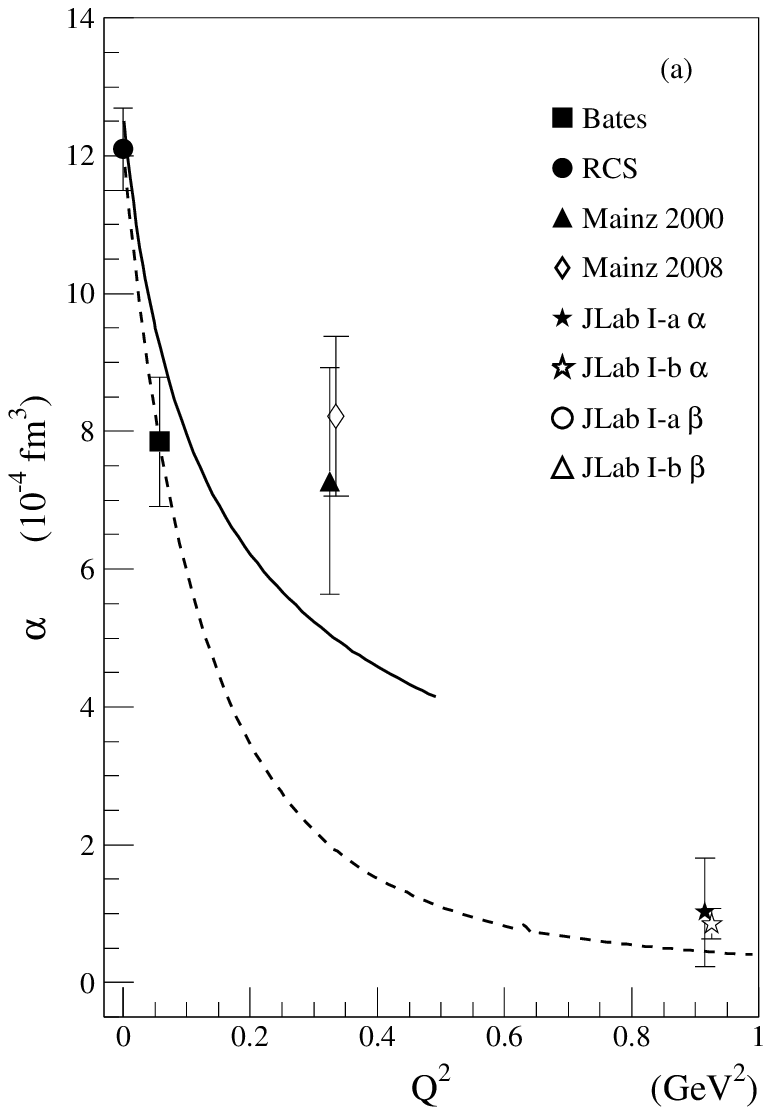,width=\textwidth}
\end{center}
\end{minipage}
\begin{minipage}[b]{0.5\textwidth}
\begin{center}
\epsfig{file=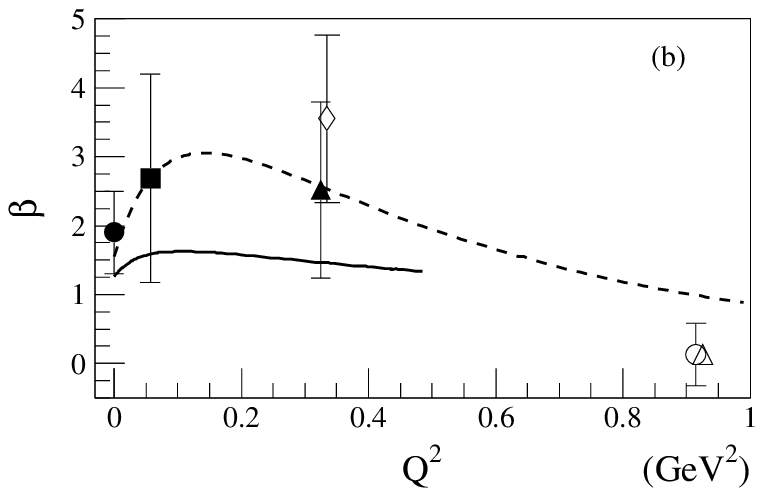,width=\textwidth}
\end{center}
\end{minipage}
\caption{Left panel: Experimental data on $\alpha_E^p(Q^2)$.
Right panel:  Experimental data on $\beta_M^p(Q^2)$.
The data are taken from Refs.\
\cite{Roche:2000ng,Laveissiere:2004nf,Bourgeois:2006js,Janssens:2008qe,Bourgeois:2011zz,Fonvieille:2012cd}.}
\label{fig:GPs}
\end{figure}

   On the experimental side, such experiments are extremely challenging since the cross section
is dominated by the Bethe-Heitler and Born processes and the structure-dependent terms involving
generalized polarizabilities represent small corrections.
   Expanding in terms of the final-state-photon center-of-mass energy, we have
\begin{equation}
d^5\sigma(pe\gamma)=d^5\sigma_{\rm BH+Born}+\omega'\left[v_{LL}(P_{LL}-P_{TT}/\epsilon)+v_{LT}P_{LT}\right]
+{\cal O}(\omega'^2),
\end{equation}
where $d^5\sigma$ is shorthand for $d^5\sigma/dk'_{\rm lab}d\Omega'_{e\,\rm lab}d\Omega_{\gamma\rm cm}$
and $v_{LL},\,v_{LT}$ are known kinematic factors.
   The generalized polarizabilities are contained in the structure functions
\begin{equation}
\begin{split}
P_{LL}-P_{TT}/\epsilon&={4m_p\over \alpha_{\rm em}}G_E^p(Q^2)\alpha_E^p(Q^2)+\text{spin GPs},\\
P_{LT}&=-{2m_p\over \alpha_{\rm em}}\sqrt{|\mathbf{q}_{cm}|^2\over Q^2}G_E^p(Q^2)\beta_M^p(Q^2)+\text{spin GPs},
\end{split}
\end{equation}
where $G_E^p(Q^2)$ is the electric form factor of the proton.
   Despite the experimental challenges, the virtual Compton scattering process has been studied at
MAMI \cite{Roche:2000ng,Janssens:2008qe}, at JLab \cite{Laveissiere:2004nf,Fonvieille:2012cd},
and at Bates \cite{Bourgeois:2006js,Bourgeois:2011zz},
with results for the generalized polarizabilities $\alpha_E^p(Q^2),\,\beta_M^p(Q^2)$ shown in
Figure \ref{fig:GPs},
   where it can be seen that there exists rough agreement with the chiral expectations.
   However, experiments with increased statistics are clearly called for.
   Finally, the generalized polarizabilities of pions and kaons have been discussed in
Refs.~\cite{L'vov:2001fz,Unkmeir:1999md,Fuchs:2000pn,Unkmeir:2001gw}.

\section{Conclusion}
   Above we have studied various aspects of hadron polarizabilities, which measure the response
of such systems to the imposition of electric and magnetizing fields, which are supplied in a
Compton scattering process.
   In the case of the nucleon, the proton and neutron electric and magnetic (scalar) polarizabilities
$\alpha_E$ and $\beta_M$, which arise at ${\cal O}(\omega^2)$ with respect to the leading Thomson amplitude,
are found experimentally to be
\begin{equation}
\alpha_E\simeq 6\beta_M\simeq 12\times 10^{-4}\,{\rm fm}^3
\end{equation}
and are in line with what is expected from heavy-baryon chiral perturbative estimates,
indicating the importance of the meson cloud in the nucleon analysis.
   In the case of the charged pion, chiral symmetry provides a strong constraint in
terms of radiative pion beta decay and ChPT makes a firm prediction beyond the
current-algebra result at the two-loop level.
   Both the experimental determination as well as the theoretical extraction from
experiment require further efforts.
   We also have studied the higher-order manifestations of polarizabilities.
   At ${\cal O}(\omega^3)$ we have discussed the four spin polarizabilities, which are just
now beginning to be measured at MAMI and potentially at HIGS.
   We also looked at polarizabilities which arise at ${\cal O}(\omega^4)$ and higher and
confronted chiral estimates with ``measured'' values from dispersion relations.
   Finally, we have looked at the generalized polarizabilities, which examine the local
polarizability distributions.
   As a general observation we conclude that the gross picture is described by the
lowest, nontrivial order in ChPT and precise measurements will provide valuable
guidance for assessing the physics beyond lowest-order dynamics.

\section*{Acknowledgment}
The authors would like to thank Vladimir Pascalutsa for providing us with the
updated version of Figure \ref{fig:alphabetav2} and Rory Miskimen
for providing Figure \ref{fig:GPs}.
   Stefan Scherer was supported
by the Deutsche Forschungsgemeinschaft (SFB 443 and 1044).

\end{document}